\documentclass[%
 reprint,
 amsmath,amssymb,
 longbibliography,
 aps,
 prx,
]{revtex4-1}

\usepackage{graphicx,subfigure}
\usepackage{amssymb, amsmath,amssymb,amsfonts}
\usepackage{amsthm,mathrsfs,amsopn}
\usepackage{dcolumn}
\usepackage{bm}
\usepackage{color}

\usepackage{rotating}
\theoremstyle{plain} 

\newcommand{\be}{\begin{equation}}
\newcommand{\ee}{\end{equation}}
\newcommand{\BE}{\begin{eqnarray}}
\newcommand{\EE}{\end{eqnarray}}
\newcommand{\BM}{\begin{multline}}
\newcommand{\EM}{\end{multline}}

\begin{document}

\title{Random walks on hypergraphs}

\author{Timoteo Carletti}
\affiliation{naXys, Namur Institute for Complex Systems, University of Namur, Belgium}

\author{Federico Battiston}%
\affiliation{Department of Network and Data Science, Central European University, Budapest 1051, Hungary}%

\author{Giulia Cencetti}
 \affiliation{Mobs Lab, Fondazione Bruno Kessler, Via Sommarive 18, 38123, Povo, TN, Italy}

\author{Duccio Fanelli}
\affiliation{Dipartimento di Fisica e Astronomia, Universit\`a di Firenze, INFN and CSDC, Via Sansone 1, 50019 Sesto Fiorentino, Firenze, Italy}

\begin{abstract}
In the last twenty years network science has proven its strength in modelling many real-world interacting systems as generic agents, the nodes, connected by pairwise edges.  Yet, in many relevant cases, interactions are not pairwise but involve larger sets of nodes, at a time. These systems are thus better described in the framework of hypergraphs, {whose} hyperedges effectively account for multi-body interactions. We hereby propose a new class of random walks defined on such higher-order structures, {and grounded on a microscopic physical model} where multi-body proximity is associated to highly probable exchanges among agents belonging to the same hyperedge. We provide an analytical characterisation of the process, deriving a general solution for the stationary distribution of the walkers. The dynamics is ultimately driven by a generalised {random walk} Laplace operator that reduces to the standard random walk Laplacian when all the hyperedges have size $2$ and are thus meant to describe pairwise couplings. We illustrate our results on synthetic models {for which we have a full control of the high-order structures}, and real-world networks where higher-order interactions are at play. {As a first application of the method}, we compare the behaviour of random walkers on hypergraphs to that of traditional random walkers on the corresponding projected networks, drawing interesting conclusions {on node rankings in collaboration networks.} {As a second application, we show how information derived from the random walk on hypergraphs can be successfully used for classification tasks involving objects with several features, each one represented by a hyperedge}. Taken together, our work contributes to unveiling the effect of higher-order interactions on diffusive processes in higher-order networks, shading light on mechanisms at the hearth of biased information spreading in complex networked systems.
\end{abstract}	
 
\maketitle

\section*{Introduction}
From social systems and the World Wide Web to economics and biology, networks define a powerful tool to describe many real-world systems~\cite{Newmanbook,Barabasibook,Latorabook}. 
Over the last twenty years of network science~\cite{AlbertBarabasi,BLMCH}, many interacting systems with different functions were shown to exhibit surprisingly similar structural properties, at different scales. Interestingly, the complex architecture of real-world networks was found to significantly interfere with the dynamical processes hosted on them,  
from social dynamics~\cite{Castellanoreview} to synchronisation~\cite{Arenasreview}. As a consequence, properly tailored dynamical processes are now routinely employed to extract information on the a priori unknown structure of the underlying graphs architectures. 

Networks materialise as pairwise interactions, represented by edges, among generic agents, the nodes: by their very first definition they are thus bound to encode binary relationships among units. However, an increasing amount of data indicates that, from biological to social systems, real-world interactions often occur among more than two nodes at a time. This phenomenon is not properly described by the traditional paradigm constrained on pairwise interactions, and highlight the need for extended notions in the realm of network theory. In recent years, an emerging stream of research has been focusing on developing higher-order network models that account for the diverse kinds of higher-order dependencies, as found in complex systems. 

Let us here observe that the current ``high-order framework'' bears some ambiguity, as it has been occasionally assumed to embrace  features which are more specifically
stemming from the interactions~\cite{LRS}, as e.g. temporal and/or memory effects~\cite{Scholtes,RELWL}, or reflect the multiplex nature of the examined system~\cite{DDGPA,kivela2014multilayer, Battiston2017challenges}. Here, the term higher-order is exclusively meant to refer to agents interacting in groups of arbitrary numerosity~\cite{BGL,BASJK,IPBL,GBMSA}, a process often modelled via simplicial complexes~\cite{DVVM,BC,PB} or hypergraphs~\cite{berge1973graphs,estrada2005complex,GZCN}, non trivial mathematical generalisation of the ordinary networks.  

Our focus is on hypergraphs, where relationships among agents are described as collections of nodes assembled in sets, called hyperedges, made by any number of nodes. Hypergraphs provide a natural representation for many higher-order real-world networks~\cite{petri2014homological, petri2018simplicial}. In social systems they can for instance be suited to describe collaboration networks, where nodes denote authors and hyperedges stand for groups of authors, who have written papers together. Alternatively,  hypergraphs can be invoked to describe face-to-face social networks where individuals can interact in groups of arbitrary sizes~\cite{patania2017shape}. In biology, hypergraphs allow to properly model bio-chemical reactions simultaneously involving more than two  species, or conveniently describe higher-order interactions among different families of proteins~\cite{BASJK}. Crucially, in all these examples, interactions among agents occur in groups of arbitrary size, and  cannot be split into disjoint pairwise interactions. Differently from simplicial complexes, a higher-order interaction described by an hypergraph (e.g. a single three body interaction) does not require the existence of all lower order interactions (e.g. the three pairwise interactions associated to the same triangle)~\cite{courtney2016generalized}. Heterogeneous hypergraphs have been sometimes studied by mapping the nodes belonging to a hyperedge into a clique of suitable size. However, the drawback of this procedure is that it eventually yields  a {\em projected} network, e.g. shown in Fig.~\ref{fig:figHnets}, where only pairwise interactions are ultimately accounted for (see Appendix~\ref{sec:projection}). 

Linear dynamics~\cite{May,Allesina,Pecora}, and specifically random walks~\cite{Rednerbook}, constitute a simple, although powerful tool to extract information on the relational structure of interacting systems. In particular, random walks on complex networks~\cite{noh2004} have been proven useful to compute centrality scores~\cite{newman2005}, finding communities~\cite{rosvall2008} and providing a taxonomy of real-world networks~\cite{nicosia2014}. In the simplest case, at each time step, a walker jumps from the node where it belongs to one of its adjacent neighbours, traveling across one of the available edges, chosen at random with uniform probability. Many variations of this fundamental process have since then been considered. These include more sophisticated dynamical implementations, which allow to targeting the walks towards nodes with given structural features~\cite{gardenes2008}, let them interact at the nodes of the network~\cite{cencetti}, investigate non-linear transition probabilities~\cite{Skardal} and crowded conditions~\cite{Asllani2018},  consider the temporal~\cite{starnini,Petit2018,Petit2019} or multilayer~\cite{dedomenico2014navigability,battiston2016} dimensions of the edges under different network topologies.

Random walks have been defined on simplicial complexes~\cite{SBHLJ}, but because of the cumbersome involved combinatorics, applications have been limited to higher-order interactions of the lowest dimensions, i.e. triangles. {Moreover, walkers are in general allowed to hop between edges or even high-order structures. This is at variance with the setting that we here aim at exploring, where hops can solely occur among nodes which join in a given high-order structure.} In parallel, also {random walks on hypergraph have been considered by assuming that all the hyperedges are made by an identical -- and constant --  number of nodes~\cite{lu2011high,HL2019}. The first random walk Laplacian defined on hypergraphs can be probably traced back to the seminal paper  by Zhou and collaborators~\cite{zhou2007learning}. Each hyperedge is endowed with an arbitrary weight, acting as a veritable bias to the walkers dynamics. As observed by the authors of ~\cite{zhou2007learning}, assigning the weights is an outstanding open problem, which deserves to be properly addressed. In this work, we will prove that a physically motivated choice for the aforementioned weights naturally emerge, when framing the problem on solid microscopic grounds.}
 
Interestingly,  more complicated nonlinear dynamics have been also recently  studied on simplicial complexes~\cite{IPBL,PB,MGA2019,JJK2019} {or in a pure multi-body frame~\cite{NML20192019}}.  Once again, however, the focus {is placed} on low-dimensional simplicial complexes (triangles). Recently,  several dynamics, including epidemic spreading~\cite{IPBL, de2019social,MGA2019} and synchronisation~\cite{Skardal}, have been shown to produce new collective behaviours when higher-order interactions are {assumed to shape the networked arrangement}.

\begin{figure}[ht]
\centering
\includegraphics[width=.4\textwidth]{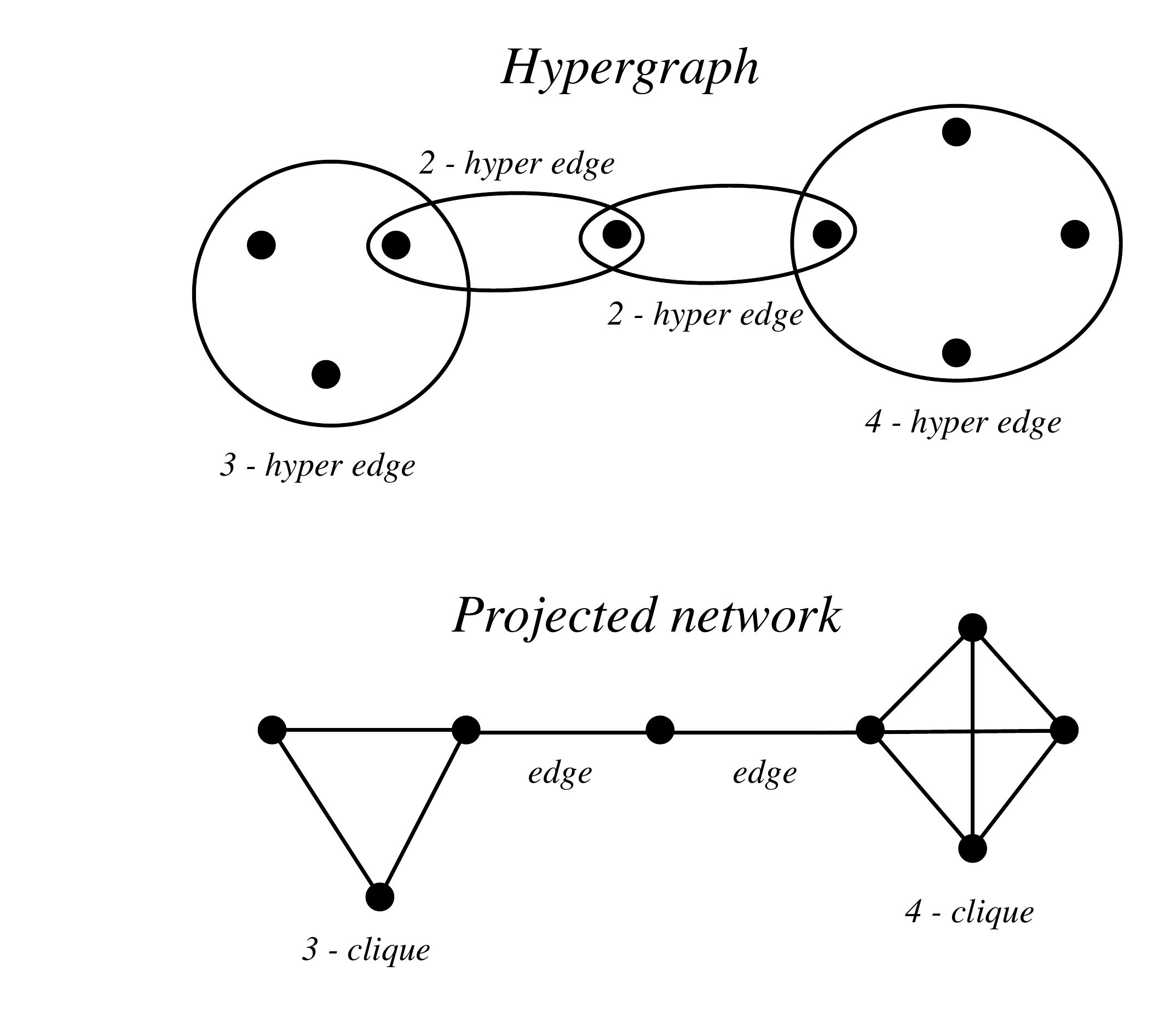}
\caption{\textbf{Hypergraph and projected network.}  Hypergraph (top) and corresponding projected network (bottom). In the projected network each hyperedge $E_\alpha$ becomes a complete clique of size $|E_\alpha |$, with thus $|E_\alpha |(|E_\alpha| -1)/2$ pairwise interactions.}
\label{fig:figHnets}
\end{figure}

Starting from this setting, {and by further elaborating on the above}, we propose {in this work} a new class of random walks, {evolving} on generic heterogeneous hypergraphs {as dictated by a plausible physical model, and} without any limitation on the sizes of the hyperedges. In this framework, multi-body proximity is associated to {highly probable} exchanges among agents belonging to the same hyperedge, and walkers mitigate their inclination to explore the system with a tendency to naturally spend more time in highly clustered cliques and communities. This feature is reminiscent of bias in information spreading, which is known to be affected by the phenomenon of echo chambers~\cite{echochambers}. Similarly to the standard random walk, at each time step a walker sitting on a node, selects a hyperedge among the ones containing the origin node, with a probability proportional to the size of the hyperedge; then the walker jumps with uniform probability onto any node contained in the selected hyperedge. In this way, higher-order interactions between a group of nodes drive the process {and the weights postulated in ~\cite{zhou2007learning} take non trivial values, as stemming from the microscopic dynamics.}

We shall in particular provide an analytical description of the process, by deriving a general formula for the stationary distribution of the walk, and show that the dynamics is driven by a generalised Laplace operator, that reduces to the standard random walk Laplacian when all hyperedges have size $2$, and the hypergraph results in a traditional network. 

As already stated, random walks can be used to rank nodes, based on the stationary occupancy probability of walkers across the network. Because of these implications, it is therefore interesting to compare the stationary distribution, as obtained within the newly introduced framework, with that displayed by standard random walkers on the corresponding projected network. Because of the tight interactions among agents belonging to the same hyperedge, the probability to find a walker on a given node is in principle different, when confronting the outcome of the two aforementioned processes. As a consequence, we expect a different order in the ranking to be obtained for the same node, depending on the dynamical process employed in the analysis. This observation opens up the way to a new definition of centrality for systems where the high-order structure is known to be relevant. In particular, we will provide a direct evidence for our claims working with co-authorship networks, as extracted from the arXiv on-line preprint server. {Our second application, to which we alluded above,  concerns a classification task which is  borrowed by ~\cite{zhou2007learning}. Indeed it is well known that one can model a dataset by resorting to networks and then make use of the associated Laplacian Eigenmaps~\cite{BN2002} to embed the data on a lower dimensional space, while hopefully preserving relevant information, in the spirit of a generalised principal component analysis. Working in the lower dimensional space allows one to cluster together objects. However, when objects to be classified share annotated features, 
the use of binary relationships, i.e. usual network, results in a dramatic loss of information. One can thus obtain a better embedding via hypergraphs and invoke the spectral characteristics of the associated Laplacian to achieve more effective clustering scores~\cite{BN2002,zhou2007learning}. By replicating the analysis in ~\cite{zhou2007learning}, we will here consider the problem of separating the animals listed in the UCI Machine Learning Depository in distinct class, e.g. mammals, birds, ..., by using at the scope a set of annotated features, e.g. tail, hair, legs and so on. Here nodes are animals and hyperedges features. We will show that the presence of high-order interactions among features as encoded via the proposed Laplacian operator, yields a very effective embedding with just a few of the most significative directions, a result which is in line with that reported in~\cite{zhou2007learning}.}
 
Summing up, we here  introduce and discuss {a novel} generalisation of the random walk picture to higher-order networked systems, {where hyperedge weights are naturally assigned and thus removing any ambiguity in their values}. Finally we hint at important exploitations of this novel dynamical framework {working along two paradigmatic directions, ranking and classification of data.}

\section*{Model}
\label{sec_model}

{\bf Incidence and hyper adjacency matrices.} 
Let us consider an hypergraph $\mathcal H(V,E)$, where $V=\{1,\dots,n\}$ is the set of $n$ nodes, $E=\{E_1,\dots,E_m\}$ the set of $m$ hyperedges, with $E_\alpha$ an unordered collection of nodes, i.e. $E_\alpha\subset V$, $\forall \alpha=1,\dots,m$. We observe that whenever $E_\alpha=\{i,j\}$, i.e. $|E_\alpha |=2$, then the hyperedge is actually a ``standard'' edge, denoting a binary interaction among nodes $i$ and $j$. An hypergraph where $|E_\alpha |=2$ $\forall \alpha$ reduces to a network.

We can define the associated {\em hyper incidence matrix} $e_{i \alpha}$, carrying the information about how nodes are shared among hyperedges, as
\begin{equation}
\label{eq:incid}
e_{i \alpha}=\begin{cases} 1 &\text{$i\in E_{\alpha}$}\\
0 & \text{otherwise}\, .
\end{cases}
\end{equation}
We note that the same matrix exists for networks. However, while in regular networks each column can have only two non zero entries, as each edge can contain two nodes only~\footnote{We do not consider here hyperedges with size $1$, because they correspond to isolated nodes, i.e. nodes that cannot take part to the examined process.}, in hypergraphs each column can display several non zeros entries (i.e. an hyperedge can contain several nodes).

Starting from the above matrix, one can construct the $n\times n$ {\em hyper adjacency matrix}, $A=ee^{T}$, whose entry $A_{ij}$ represents the number of hyperedges containing both nodes $i$ and $j$. We note that often the adjacency matrix is defined by setting to $0$ the main diagonal. Let us also define the $m\times m$ {\em hyperedges matrix}, $C=e^{T}e$, whose entry $C_{\alpha \beta}$ counts the number of nodes in $E_{\alpha}\cap E_{\beta}$. {Observe that in the literature the number of nodes in a given hyperedge, $C_{\alpha\alpha}$, is often called the degree of the hyperedge, while the node degree stands for the number of hyperedges containing the node, $\sum_\alpha e_{i\alpha}e_{i\alpha}$.}

{\bf Transition probability.} 
To describe a random walk process, we need to define the transition probability to pass from a state, hereby represented by the node on which the walker belongs to, to any other state, compatible with the former, in one time step. In the case of simple unbiased random walks on networks, one assumes the walker to take with equal probability any link emerging from the node that is initially occupied. Hence, the transition probability can be readily computed as $A_{ij}/k_i$, where $k_i=\sum_j A_{ij}$ is the degree of the origin node. When dealing with hypergraphs, by choosing with uniform probability any of the neighbouring nodes, namely all the nodes belonging to hyperedges connected with the origin node, is not a sensible choice. In this way, in fact, the real structure of the systems is not incorporated into the dynamical picture. On the contrary, nodes belonging to the same hyperedge exhibit a higher-order interaction and we consequently assume that spreading among them is {more probable} than with nodes associated to other hyperedges; because of this the information can thus spend long periods inside the same hyperedge. For instance, a gossip can spread faster because of group interaction among individuals, than as follows successive binary encounters; similarly, ideas can circulate more effectively among collaborators, the coauthors of a joined publication, as compared to the setting where exchanges in pairs are solely allowed for. To compute the transition probability to jump from $i$ to $j$, we count the number of nodes, excluding $i$ itself, belonging to the same hyperedge of $i$ and $j$. Recalling the definition of the matrix $C$, this can be written as
\begin{equation}
\label{eq:khij}
k^{H}_{ij}=\sum_{\alpha}(C_{\alpha \alpha}-1)e_{i\alpha}e_{j\alpha}=(e\hat{C}e^T)_{ij}-A_{ij}\quad\forall i\neq j\, ,
\end{equation}
where $\hat{C}$ is a matrix whose diagonal coincides with that of $C$ and it is zero otherwise (see Appendix~\ref{sec:Tp}).
By normalising so as to impose a uniform choice among the connected hyperedges, we get the following expression for the transition probabilities:
\begin{equation}
\label{eq:Tij4}
T_{ij}=\frac{(e\hat{C}e^T)_{ij}-A_{ij}}{\sum_\ell k^{H}_{i\ell}}=\frac{(e\hat{C}e^T)_{ij}-A_{ij}}{\sum_\ell (e\hat{C}e^T)_{i\ell}-k^H_i}\, ,
\end{equation}
where $k^H_i=\sum_\ell A_{i\ell}$ is the hyperdegree of the node $i$, a synthetic measure reminiscent of the node degree, which takes 
into account both the number and the size of hyperedges $i$ in which $i$ is involved.

When the hypergraph is a network, all hyperedges have $2$ nodes. Hence
\begin{equation}
(e\hat{C}e^T)_{ij}=\sum_{\alpha}C_{\alpha \alpha}e_{i\alpha}e_{j\alpha}=2\sum_{\alpha}e_{i\alpha}e_{j\alpha}=2A_{ij}\, ,
\end{equation}
and Eq.~\eqref{eq:Tij4} reduces to the standard transition probability for random walk on networks
\begin{equation}
T_{ij}=\frac{2A_{ij}-A_{ij}}{2k^H_i-k^H_i}=\frac{A_{ij}}{k_i}\, ,
\end{equation}
where we used the fact that, under this assumption, $k^H_i=k_i$.

{\bf Stationary solution.}
Having computed the transition probabilities, we can proceed further by formulating the dynamical equation which rules the temporal evolution of the probability 
${\bf p} (t)= (p_1(t), \ldots, p_n(t))$ of finding the walker on a given node after $t>0$ steps. The process is governed by the following  equation: 
\begin{equation}
\label{eq:trans}
{p}_i(t+1)=\sum_j p_j(t)T_{ji}\, ,
\end{equation}
where the right hand side term combines the probability to be in any node $j$ at time $t$ and the probability to perform a jump towards the target node $i$, during the next time of iteration.  As $\sum_jT_{ij}=1$ for all $i$, the stationary probability distribution, ${\bf p}^{(\infty)}$, is thus the left eigenvector associated with the eigenvalue $\lambda_1=1$ of $\mathbf{T}$.

Given $\mathbf{T}$, it is possible to obtain an exact analytical solution for the stationary state ${\bf p}^{(\infty)}$ which encapsulates the higher-order structure of the system. Indeed, a straightforward computation (see Appendix~\ref{sec:statsol}), yields: 
\begin{equation}
p_j^{(\infty)}=\frac{\sum_\ell (e\hat{C}e^T)_{j\ell}-k^H_j}{\sum_{m\ell}\left[ (e\hat{C}e^T)_{m\ell}-k^H_m\right]}\,,
\label{eq:statnorm}
\end{equation}
for all $j=1,\dots,n$. In the case the hypergraph is indeed a network, we recover the well known expression $q_j^{(\infty)}= k_j/\sum_l k_l$ for the stationary solution of the walk. {Let us observe that 
\begin{equation}
\label{eq:rwLap}
L_{ij}=  \delta_{ij}-T_{ij} = \delta_{ij}-\frac{k^H_{ij}}{\sum_\ell k^H_{i\ell}}\, ,
\end{equation}
is a new {{\em random walk}} Laplacian that generalises the random walk one for networks. Moreover the former reduces to the latter in the case $|E_\alpha|=2$ for all $\alpha$.}

We observe that the formalism readily extends to the case of continuous-time random walks, where the evolution of the probability is given by 
\begin{equation*}
\dot{p}_i(t)=\sum_j p_j(t)T_{ji} - \sum_j p_iT_{ij}\, .
\end{equation*}
Similarly to the case of networks, as $\sum_j T_{ij}=1$, it is possible to rewrite the latter as
\begin{equation*}
\dot{p}_i=\sum_j p_j(T_{ji} - \delta_{ij})=-\sum_j p_jL_{ji}\, ,
\end{equation*}
{where $\mathbf{L}$ is the above defined Laplace matrix.} In the following, {for a sake of for the sake of definiteness we limit our analysis to} explore the properties of the discrete-time random walks on synthetic and real-world hypergraphs, {leaving the continuous time case to a further work}.

{Denote by $\mathbf{D}$, the diagonal matrix with entries $d_i^H=\sum_j k_{ij}^H$ and by $\mathbf{K}^H$, the matrix characterised by elements $k_{ij}^H$. We can introduce the {\em symmetric} Laplacian $\mathbf{L}^{sym}$ as:
\begin{equation*}
\mathbf{L}^{sym}=\mathbf{I}-\mathbf{D}^{-1/2}\mathbf{K}^H\mathbf{D}^{-1/2}\, ,
\end{equation*}
which is well defined since $k_{ij}^H\geq 0$. $\mathbf{L}^{sym}$ is similar to the operator introduced via relation $(\ref{eq:rwLap})$, indeed $\mathbf{L}=\mathbf{D}^{-1/2}\mathbf{L}^{sym}\mathbf{D}^{1/2}$.  The newly introduced operator $\mathbf{L}$ is
hence a properly defined Laplacian: it is in fact non-negative definite, it displays real eigenvalues and the smallest eigenvalue is identically equal to zero, as it readily follows by virtue of the proven similarity with $\mathbf{L}^{sym}$.}

{Before turning to discussing the applications, we will briefly draw a comparison with the setting proposed by Zhou~\cite{zhou2007learning} and show how this materialises in a natural solution for the problem of weights determination. The Laplacian operator $\mathbf{L}^{z}$ introduced in ~\cite{zhou2007learning} can be cast in the form:}
\begin{equation}
\label{eq:Lzhou}
L^{z}_{ij}=\delta_{ij}-\sum_\alpha \frac{w_\alpha}{W_i C_{\alpha\alpha}} e_{i\alpha} e_{j\alpha}\, ,
\end{equation}
{where $w_\alpha$ identifies the undetermined weight of the hyperedge $E_\alpha$, $W_i=\sum_\alpha w_\alpha e_{i\alpha}$ is the total weight of the hyperedges containing the node $i$, i.e. weighted node degree, and $C_{\alpha\alpha}$ stands for the number of nodes in the hyperedge $E_\alpha$. A simple calculation, as detailed in the following, shows that operator  $\mathbf{L}$ can be eventually recovered from $\mathbf{L}^{z}$ by imposing the non trivial weights $w_\alpha=C_{\alpha\alpha}(C_{\alpha\alpha}-1)$. In fact:}

\begin{widetext}
\begin{eqnarray*}
L^{z}_{ij}&=&\delta_{ij}-\sum_\alpha \frac{C_{\alpha\alpha}(C_{\alpha\alpha}-1)}{C_{\alpha\alpha}\sum_\beta C_{\beta\beta}(C_{\beta\beta}-1) e_{i\beta} } e_{i\alpha} e_{j\alpha}=\delta_{ij}-\sum_\alpha \frac{(C_{\alpha\alpha}-1)}{\sum_\beta C_{\beta\beta}(C_{\beta\beta}-1) e_{i\beta} } e_{i\alpha} e_{j\alpha}\\
&=&\delta_{ij}-\frac{k_{ij}^H}{\sum_\beta \sum_\ell e_{\ell\beta}(C_{\beta\beta}-1) e_{i\beta}}=\delta_{ij}-\frac{k_{ij}^H}{ \sum_\ell k_{i\ell}^H} = L_{ij}\, ,
\end{eqnarray*}
\end{widetext}
{where used has been made of definition~\eqref{eq:khij} for $k_{ij}^H$ and the fact that $C_{\beta\beta}=\sum_\ell e_{\ell \beta}$. As anticipated, a natural choice for the weights as postulated in~\cite{zhou2007learning} can be envisaged, which follows a sensible microscopic modelling of the random walk dynamics.}

{By invoking  Theorem 4 in~\cite{chitraraphael2019}, we can finally conclude that our process is equivalent to a random walk on a weighted projected network, where the weights of the link $ij$ is given by $k_{ij}^H$, that is the weights scale extensively with the region of influence of the nodes, namely the size of the hyperedge they belong to. It is indeed quite remarkable that a properly weighted binary network encapsulates the higher order information, as stemming for the corresponding hypergraph representation.
Observe that authors in~\cite{chitraraphael2019} also consider an extension of the Zhou et al. model, where nodes bear a given weight, tuned so as to reflect the hyperedge characteristics. Again, the introduced weights are abstract quantities, and do not reflect a physically motivated choice.}

\section*{Results}
\label{sec_results}

\begin{figure*}[ht]
\centering
\includegraphics[width=.8\textwidth]{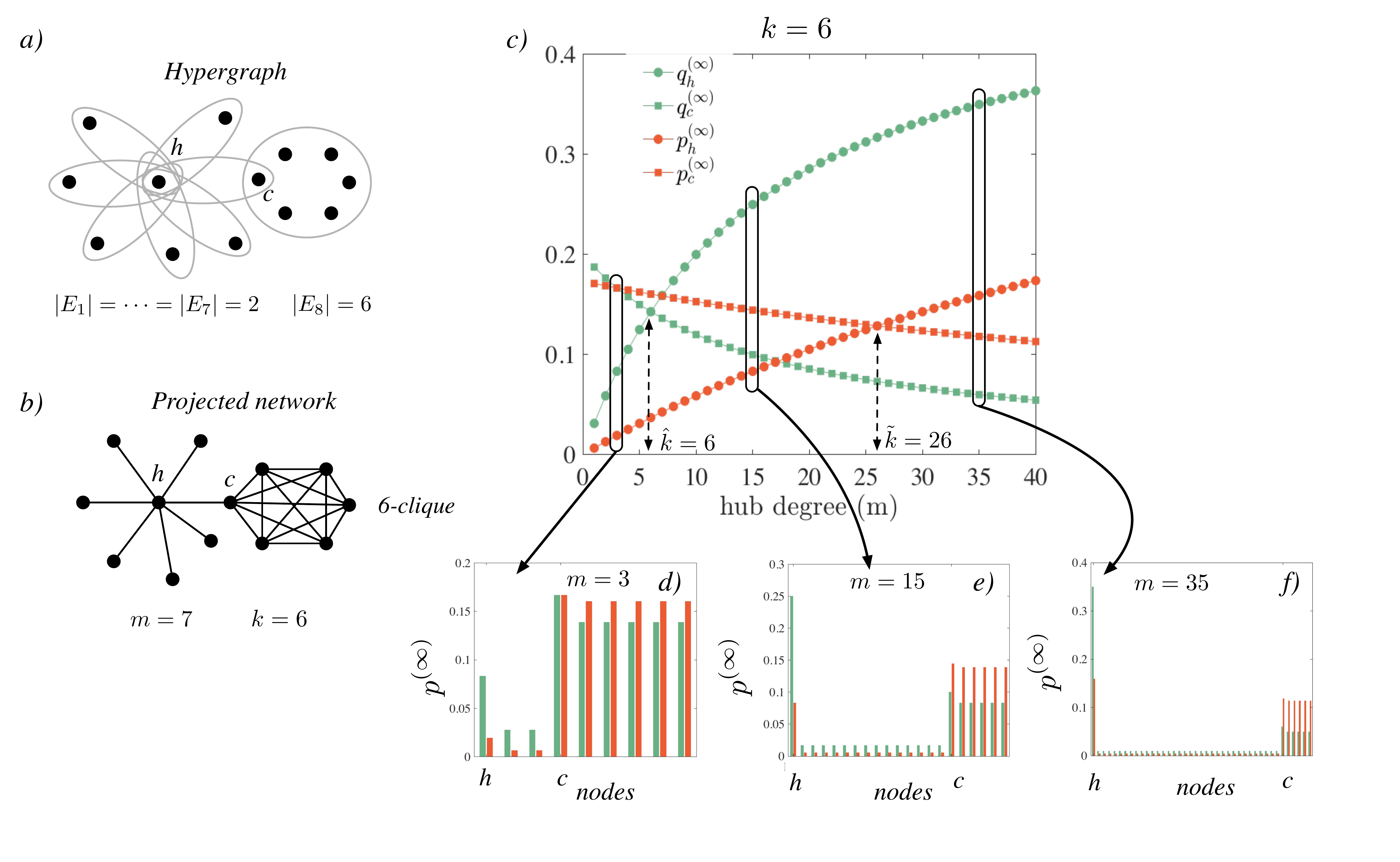}
\caption{\textbf{The $(m,k)$-star-clique network}. Panel a: hypergraph made by $m+k=13$ nodes, divided into $m=7$ hyperedges of size $2$ and one large hyperedge of size $k=6$. The node $h$ belongs to all the $2$-hyperedges, while the node $c$ belongs to one $2$-hyperedge and to the $6$-hyperedge. Panel b: the projected network where hyperedges are mapped into complete cliques, the $6$-hyperedge becomes thus a $6$-clique. Panel c: we show the dependence on $m$ of the asymptotic probability of finding the walker on the node $h$ (circle) or on the node $c$ (square), in the projected network (green symbols) and in the hypergraph (orange symbols). Panels d, e and f: we report the asymptotic probabilities $q_i^{(\infty)}$ and $p_i^{(\infty)}$ for three values of $m$: $m=3<\hat{k}$, $\hat{k}<m=15<\tilde{k}$ and $\tilde{k}<m=35$, where $\hat{k}=6$ and $\tilde{k}=26$.}
\label{fig:figHnetsEx}
\end{figure*}

Since the Page-Rank~\cite{ilprints361,ilprints422}, random walks on networks are routinely applied to compute centrality scores~\cite{newman2005}. Indeed they can be used to rank nodes according to the probability to be visited by the walker, the larger the latter the more ``important''/``central'' the node. In this section, we show that high-order interactions can strongly modify the ranking, as resulting from a random walk process on hypergraphs, with respect to the homologous estimate as computed for the corresponding projected network. This fact can thus bear relevant implications for ranking real data, stemming from a dynamical process which is better explained in terms of hypergraphs. In this case, in fact, the applications of ranking tools tailored to pairwise interactions might produce misleading results (see Appendices~\ref{sec:statsol} and~\ref{sec:arxiv}).

To illustrate the effect of a non-trivial higher-order structure, we consider a simple hypergraph made by $m$ hyperedges of size $2$ all intersecting in a common node, $h$; a different node, say $c$, belongs to one of such $2$-hyperedges and to a hyperedge of size $k$ (see Fig.~\ref{fig:figHnetsEx} panel $a)$ for the case $m=7$ and $k=6$).

The random walk on the projected network will rank nodes according to their degree, i.e. $q_i^{(\infty)}\sim k_i$. Hence for $m>k$, the node $h$ with $k_h=m$, is ranked first, followed by the $c$ node, $k_c=k$, and all other ones (see green curves in panel c Fig.~\ref{fig:figHnetsEx}). In contrast, the random walk on the hypergraph ranks nodes taking into account higher-order relations. Since from Eq.~\eqref{eq:statnorm} we get $p^{(\infty)}_{h}\sim m$ and $p^{(\infty)}_{c}\sim 1+(k-1)^2$, thus $h$ is the top node as long as $m >1+(k-1)^2$ (see orange curves in panel c Fig.~\ref{fig:figHnetsEx}). In conclusion, for a fixed size of the hyperedge $k$, if the ``hub'' node is too small (see panel d), $m<\hat{k}=k+1$, or the hub is very large (see panel f), $m>\tilde{k}=1+(k-1)^2$, then both processes will rank nodes in the same way. However, there exists a range of intermediate values, $\hat{k}<m<\tilde{k}$, for which the top ranked node on the hypergraph is the $c$ node while the random walk on the projected network returns the $h$ node as top rank (see panel e). This phenomenon of ranking inversion will be further discussed in Appendix~\ref{sec:statsol}. In the aim of maximising the {probability of} occupancy of a given node, it is preferable for this latter to be connected to nodes organised into few large hyperedges, than to many parcelled units.

To further characterise the impact of the high-order interactions on diffusion on larger systems, we consider a second {synthetic} model where all nodes have the same number of neighbours, which are arranged in a tuneable number of triangles, i.e. hyperedges of size $|E_\alpha|=3$. The model interpolates between the case where the number of triangles is zero, $f=0$, meaning that all interactions involve simple pairs, and the case where there are no pairwise interactions but only $3$-body ones, $f=1$. More precisely, we start with a $1$D regular lattice where  nodes are connected to  $4$ neighbours ($2$ on the left and $2$ on the right). Each nodes has hence degree $4$ and takes part to $2$ distinct triangles, i.e. hyperedges with size $3$, and $f=1$. Then with probability $p$ we iteratively swap the ending points of the links with a ``criss-cross'' rewire, {i.e. preserving the nodes degree}, progressively eliminating $3$-hyperedges, hence triangles. In the limit of high rewire triangles have a negligible probability to be formed, and one eventually obtains a regular random graph with degree $k=4$. In the process, we control that no hyperedge of size greater than $3$ is created, so that competition is only between $2$-body and $3$-body interactions.

\begin{figure}[ht]
\centering
\includegraphics[width=.35\textwidth]{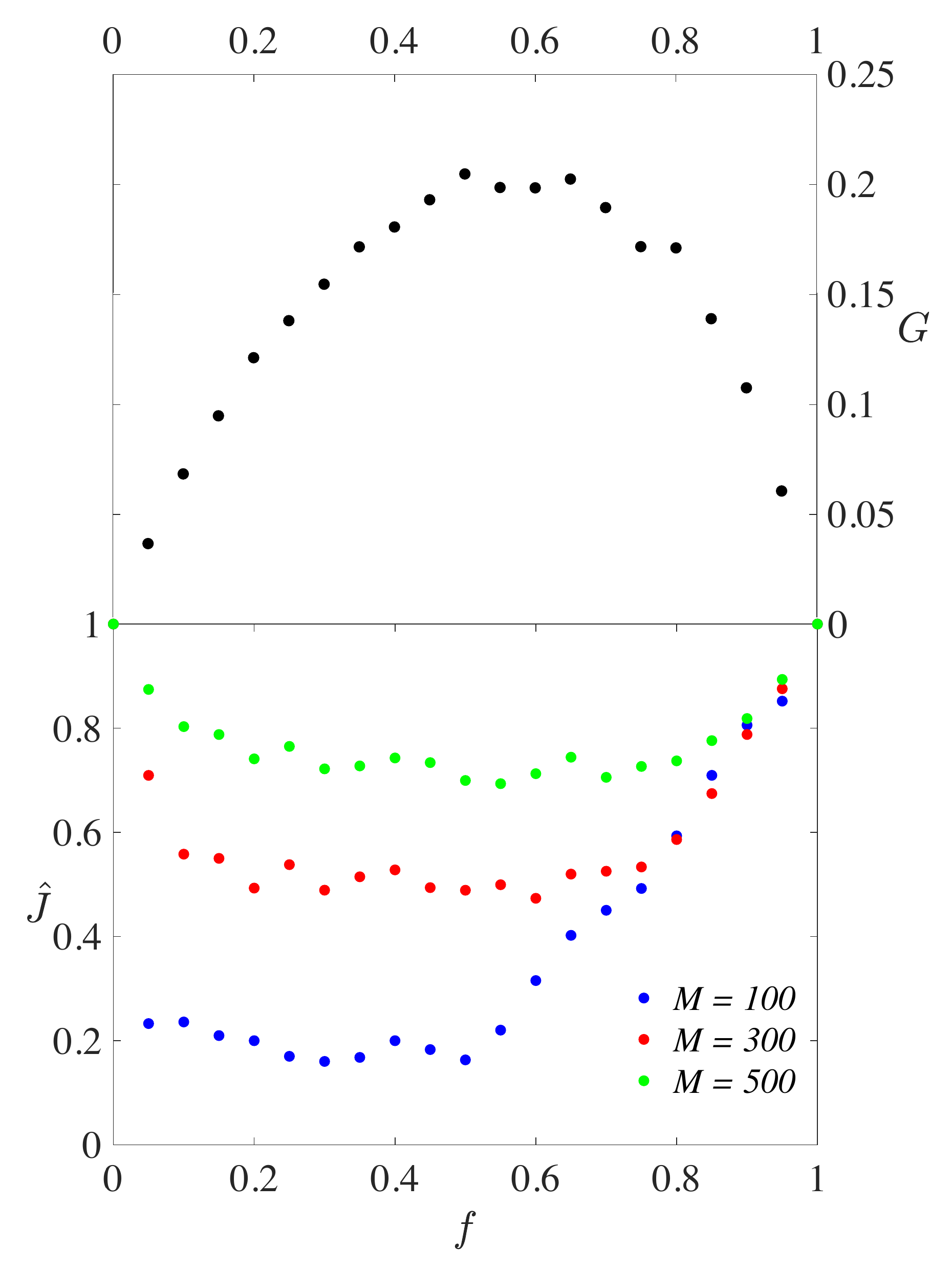}
\caption{\textbf{Impact of the $3$-body interaction on the asymptotic solution of the random walk on the hypergraph.} The top panel reports the Gini coefficient for the stationary state of the random walk on hypergraphs as a function of the fraction of hyperedges of size three, $f$. Recall that the model does not allow for hyperedges of size larger than $3$. The bottom panel shows for the same networks the modified Jaccard index in order to compare the rankings of nodes for the hypergraph and the projected network. Different colours (blue, red and green), correspond to different numbers of nodes chosen for the comparison, i.e. the top $100$, the top $300$ and all the $500$ nodes, respectively.}
\label{fig:Triangles}
\end{figure}

As the degree sequence is unchanged throughout this process and every node shares the same number of links, the asymptotic distribution of walkers on the projected network is uniform and given by $q_i=1/N$ for all $i$, where $N$ is the number of nodes, set to $500$ in the example below, no matter the value of $f$. This is also the case for the random walk on hypergraph, in the two limiting cases $f=0$ and $f=1$; indeed in the former case the hypergraph and the projected network do coincide because all the hyperedges have size $2$. In the latter setting, all nodes are involved in the same number of higher-order interactions and thus they are all equivalent. However, for the walk on hypergraphs the stationary state changes at the intermediate stages of $f$. 
In order to quantify the heterogeneity of the stationary state we rely on the Gini coefficient, which is defined as the average absolute difference between all pairs of elements in the vector $p$, divided by the average:
\begin{equation}
G(p) = \frac{\sum_{i=1}^N \sum_{j=1}^N |p_i-p_j|}{2N\sum_{i=1}^N p_i}.
\end{equation}
The Gini coefficient for the stationary state of random walk on the above described hypergraph is reported in top panel of Fig.~\ref{fig:Triangles}. 
For the limiting values $f=0$ and $f=1$ the stationary state on the hypergraph coincides with the one on the projected network {and the Gini index is $0$ being the asymptotic solution homogeneous}. However, high-order structures arising for intermediate values of the fraction of triangles induce  a heterogeneity in the occupation of the different nodes at equilibrium, which is thus different from the one obtained for the associated projected network.

A standard metric to compare lists is the Jaccard index, a measure of the fraction of elements that are common between two lists with respect to the total number of involved elements, $J(A,B)=|A\cap B|/|A\cup B|$. As the Jaccard index does not take into account the order of the elements as appearing in the  two confronted lists, we compare the rankings of the two stationary distributions by means of a modified Jaccard index, $\hat{J}$, recently introduced in~\cite{GargiuloEtAl2016}. Here differences at the top of the ranking induce a stronger change, than differences associated to the lower ranked elements. Let us observe also that the Jaccard index is unable to detect a permutation in the order of the elements in a list, while the modified one does.  In the bottom panel of Fig.~\ref{fig:Triangles}, we show the average modified Jaccard index, $\hat J$, for the $M$-top ranking, $M=100,300,500$,  as a function of the fraction $f$ of $3$-hyperedges existing in the system. The results are in agreement with the ones obtained via the Gini coefficient; {for $f=0$ and $f=1$ the rankings do coincide and thus $\hat J$ achieves is maximum value, i.e. $1$, while for intermediate values of $f$ the index $\hat J$ drops down reflecting differences among the rankings}. Moreover, we can appreciate the presence of a large turnover in the top lists: indeed $\hat{J}$ associated to small $M$, i.e. comparing relatively few nodes in the top list, are much smaller than that for large $M$, i.e. longer lists.

\begin{figure}[ht]
\centering
\includegraphics[width=.35\textwidth]{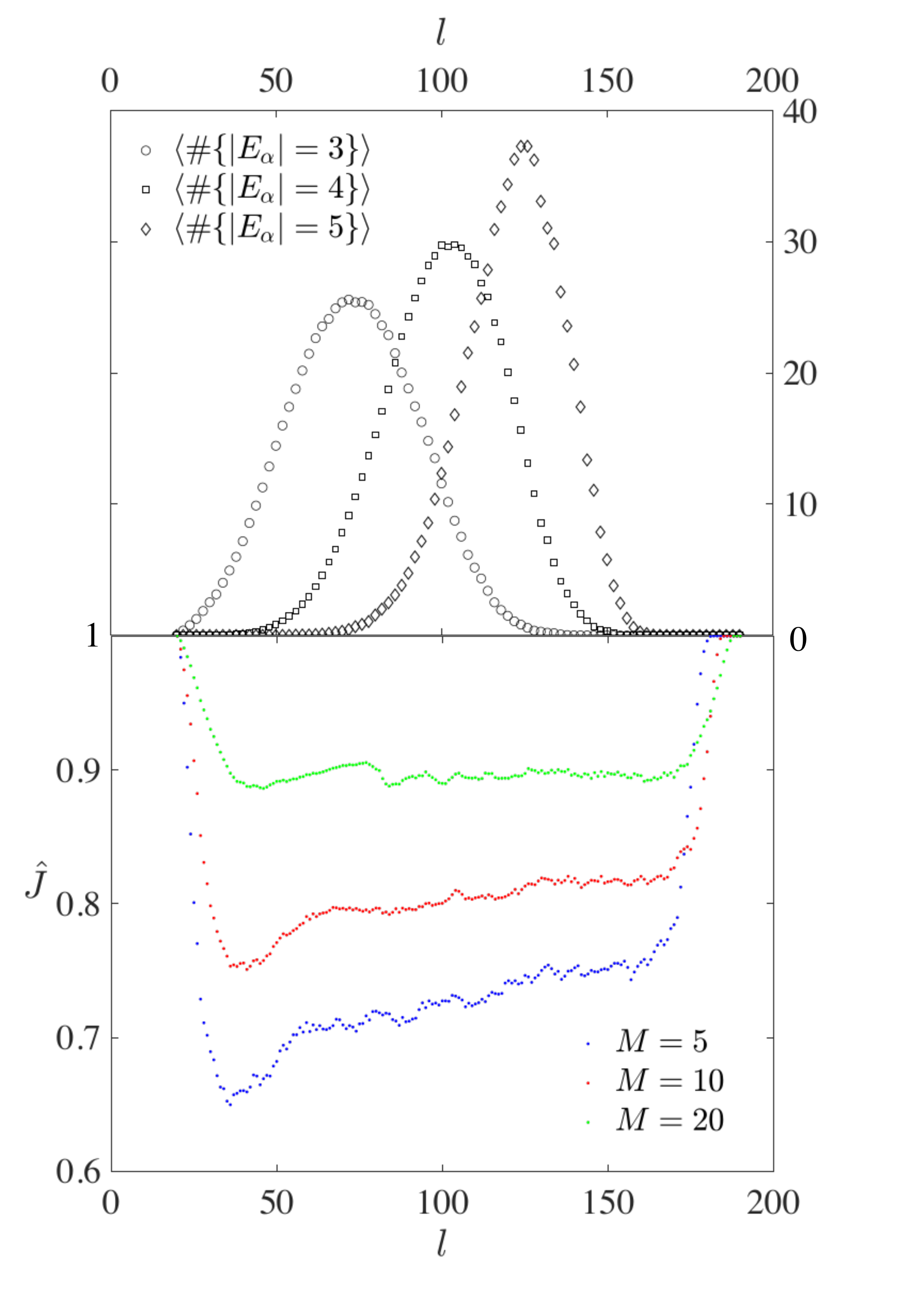}
\caption{\textbf{Impact of high-order structures on the asymptotic distribution of walkers for the random walk on the hypergraph and on the projected network.} Using the algorithm presented in the text, by iteratively adding links we create hypergraphs that interpolate from a regular $1D$ ring (where $N=20$ nodes are connected each one with its two neighbours) to a complete graph. We then perform the random walk process on respectively the hypergraphs and the associated projected network and compare the resulting ranking (the top $5$ blue, the top $10$ red and the top $20$ green, i.e. the whole set of nodes) using $\hat{J}$ (bottom panel). For a small number of available links, $l$, the hypergraphs does not present many hyperedges and thus the ranking are very close, $\hat{J}\sim 1$. As $l$ starts to increase, few hyperedges of size $3$ are created (see circles in the top panel) and the rankings estimated with the two alternative methods deviate, the values of $\hat{J}$ dropping in turn. However, as $l$ increases even more, larger high-order structures, e.g. $4$ and $5$ hyperedges, emerge (see square and diamond symbols in the top panel) and $\hat{J}$ steadily increases. For a large ensemble of added links, $l\gtrsim 170$,  the rankings become  similar and $\hat{J}\sim 1$.}
\label{fig:RingLinks}
\end{figure}

To take one step forward, we consider a {synthetic} model where high-order structures are not limited to $3$-body but larger hyperedges are allowed for. We thus build a third model which interpolates from a $1$D ring to a fully connected network. More precisely, we start from a $1$D ring where all the nodes have degree $2$, and then progressively increase its density as measured by the total number of links, $l$, until  the process terminates with a complete network, corresponding to a hypergraph with a single hyperedge containing all the nodes. Links are added at random avoiding self-loops and multiple links. We note that differently from the previous case, at intermediate values of $l$, this model presents a much wider variety in the size of the hyperedges (or cliques in the projected network), which are not anymore limited to $2$-body and $3$-body interactions. For this reason, the structure of the ranking difference is definitely more complex and rich than what one could eventually guess by just looking at the number of $3$-hyperedges, $4$-hyperedges or $5$-hyperedges (see Fig.~\ref{fig:RingLinks}).

In the initial configuration of a $1$D ring, the stationary solutions of the hypergraph and the projected network coincide, because of the absence of higher-order interactions. Similarly, they are also equivalent in the opposite limit, i.e. when the fully connected network is generated. For intermediate number of added links, the two processes result instead in different rankings. In Fig.~\ref{fig:RingLinks}, we report $\hat J$ as a function of the total number of links $l$, to compare the $M$-top rankings, as obtained by using the random walk on the hypergraph and on the projected network, respectively. We reports in particular results for three values, $M=5,10,20$. The behaviour of the threes curves is qualitatively similar. Indeed they all reach the value $1$, i.e. perfect matching of the respective rankings for $l=20$ (initial $1$D ring). Then, even the addition of just few links makes the rankings to change abruptly and $\hat{J}$ consequently drops to low values. This is associated with the creation of small hyperedges with size equal to $3$ ({see bottom panel of Fig.~\ref{fig:RingLinks}}). Adding more links reduces the differences, namely $\hat{J}$ increases, up to $l=190$ (complete network) where again the rankings do coincide and the index equals $1$. This is associated with the birth of larger hyperedges. Let us remark that $\hat{J}$ for $M=5$ is much smaller than the same quantity computed with $M=10$ (rank half of the nodes) and $M=20$ (rank all the nodes) meaning that there is a strong turnover in the top positions. {The heterogeneity in the stationary solutions of this model, as well as the star-clique example, is further investigated in the Appendix~\ref{sec:hetero} where the corresponding Gini coefficients are shown.}

\section*{Applications}
\label{sec_app}

\subsection*{Node ranking}
\label{sec_app_arxiv}

{In the previous section we have shown that hypergraph and the projected network can exhibit different stationary solutions because of  ranking inversion  (see Appendix~\ref{sec:statsol}). We thus decided to analyse the impact of this observation in real networks of scientific collaborations, in our opinion one of the most representative examples of high-order structures in human interactions. The analysed data have been gathered from the arXiv database (see Appendix~\ref{sec:arxiv} for more details). Human collaborations are often schematised as resorting to pairwise interaction, a working ansatz which amounts to  ignoring the organisation in teams. At variance, we have instead built a hypergraph where researchers (i.e nodes) co-authoring an article are part of the same hyperedge.}

\begin{figure*}[ht]
\centering
\includegraphics[width=.9\textwidth]{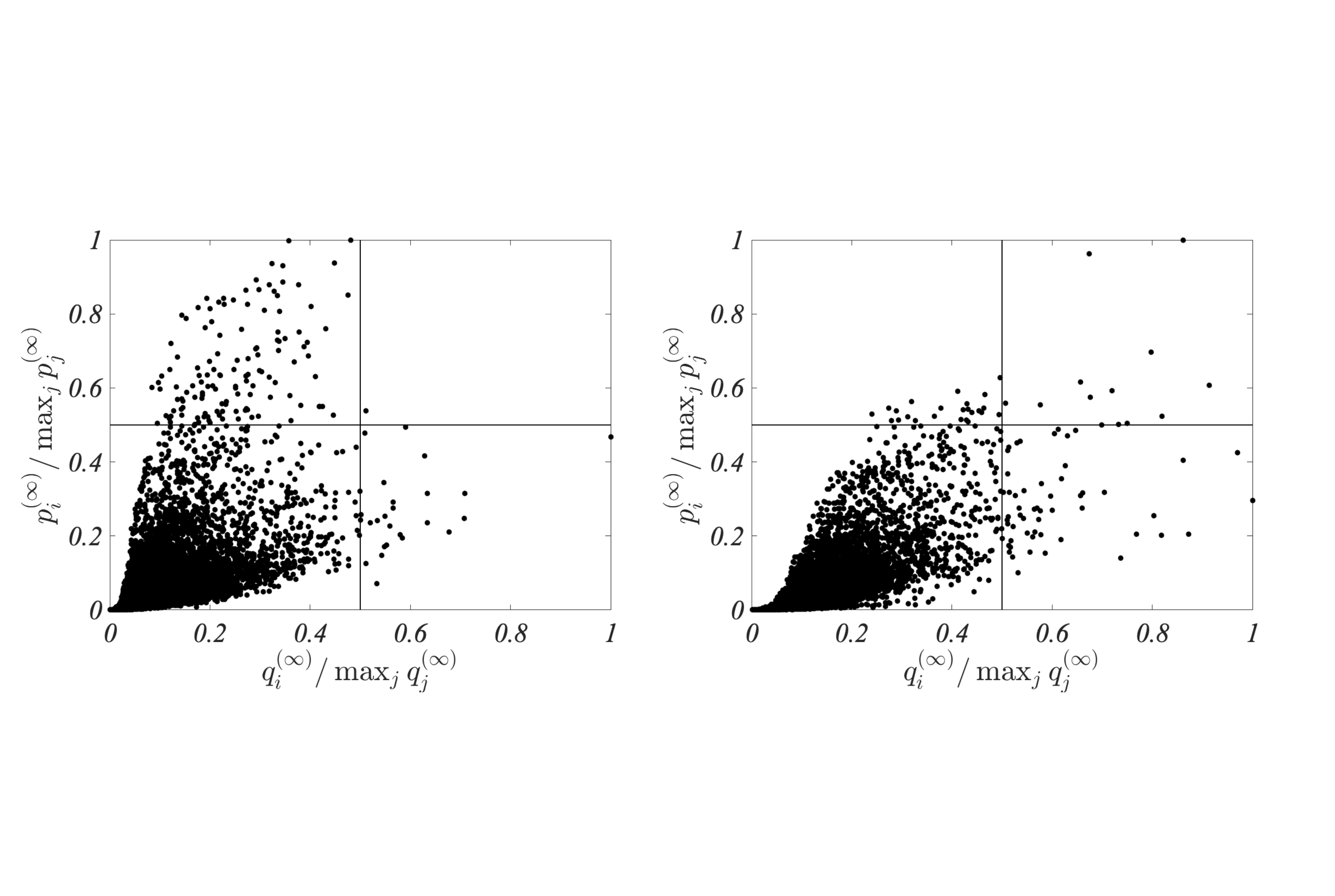}\vspace{-2cm}
\caption{{\textbf{Comparing the rankings in the arXiv community: the case of Astro and Physics}. We report the scatter plot of the normalised rankings obtained with the RW on network, $q_i^{(\infty)}$, and the one computed using the random walk on hypergraphs, $p_i^{(\infty)}$ for the arXiv-astro (left panel) and the arXiv-physics (right panel).}}
\label{fig:arxivQPAstroPhys}
\end{figure*}

{We have then determined the largest connected component of the hypergraph and that of the projected network, considered maximal and unique hyperedges (to have a fair comparison with the cliques) and computed: (i) the stationary distribution $\mathbf{p}^{(\infty)}$ for the random walk on the associated hypergraph, (ii) the stationary distribution $\mathbf{q}^{(\infty)}$ for the random walk on the corresponding projected network. We then normalise the computed stationary probabilities by their relative maximum so as to favour a comparative visualisation. In Fig.~\ref{fig:arxivQPAstroPhys} we plot 
$p^{(\infty)}_i/\max_j p_j^{(\infty)}$ vs $q^{(\infty)}_i/\max_j q_j^{(\infty)}$ for the case arXiv-astro and arXiv-physics. In Fig.~\ref{fig:arxivQP} the same comparison is drawn for the complete arXiv dataset.}

{Author are ranked differently, according to the two criteria, the one based on hypergraphs being more sensitive to the organisation in groups. If the computed rankings were (almost) the same, the data would (almost) lie on the main diagonal; deviation from this, results in novel information conveyed by the random walk on the hypergraph. The unitary square in the plane $(q^{(\infty)}_i, p^{(\infty)}_i)$ can be divided into four smaller squares (see Fig.~\ref{fig:arxivQPAstroPhys}). The majority of the authors lies in the bottom left square, $[0,1/2]\times [0,1/2]$: these authors have therefore written a few papers with a small number of co-authors. Three other regions can be however identified which roughly correspond to the bounded squares: $[1/2,1]\times [0,1/2]$ (bottom right), $[0,1/2]\times [1/2,1]$ (top left) and $[1/2,1]\times [1/2,1]$ (top right). Authors in the top right square are top ranked in both processes: they have hence written a large number of papers with different collaborators (large degree), but they have also contributed to a relevant number of papers with many co-authors, i.e. large hyperedge size. Scholars in the bottom right square are better ranked by the random walk on the network; this means that they have written several papers but with a small number of co-authors {(see e.g. right panel corresponding to physics in Fig.~\ref{fig:arxivQPAstroPhys})}. Finally, researchers in the top left square manifest a complementary attitude: they have participated to a small number of papers, but written by many authors {(see e.g. left panel corresponding to astro in Fig.~\ref{fig:arxivQPAstroPhys}).}}

\begin{figure}[ht]
\centering
\includegraphics[width=.45\textwidth]{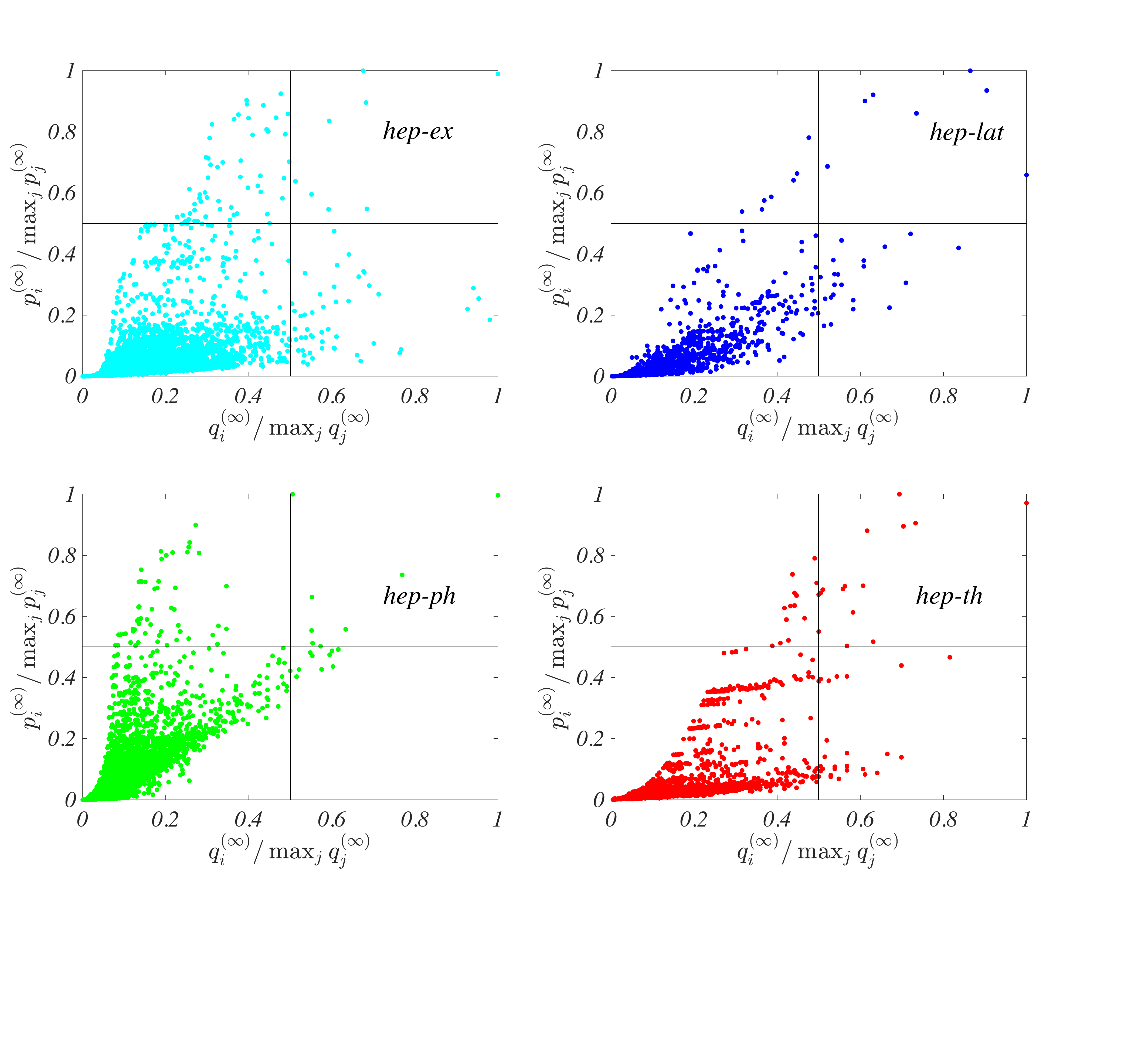}
\vspace{-1.5cm}
\caption{{\textbf{Publication habits arXiv-hep}. We report the scatter plot of the normalised rankings obtained with the random walk on the projected network, $q_i^{(\infty)}$, and the one computed using the RW on hypergraphs, $p_i^{(\infty)}$ for the four subdomains of the arXiv-hep domain.}}
\label{fig:arxivQPhep}
\end{figure}

{As a further consideration, we can bring to the fore different  ``habits'' of publication and writing papers that authors exhibit in each domain, despite the distribution of node degrees, i.e. number of different collaborators per author, and of hyperedges size, i.e. number of co-authors in papers, shows a quite similar shape across domains, as e.g. broad tails (see annexed supplementary information).
This is particularly relevant for the High Energy Particle (hep) archive, one among the oldest ones and divided into four subcategories, experimental (ex), lattice (lat), phenomenology (ph) and theory (th) (see Fig.~\ref{fig:arxivQPhep}). Indeed hep-ex and hep-ph populates mainly the top right square, while hep-lat and hep-th are more present in the top right and bottom right squares. Researchers belonging to the former community tend therefore to  write several papers with many co-authors, while those associated to the latter have papers with many different collaborators, each one co-authored by a small number of scholars. This is also confirmed by the largest degree found in the four subcategories (see Table~\ref{tab:tablearxiv}) which is as large as $\sim 1200$ for hep-ex and hep-ph, while it is almost $4$ times smaller for hep-lat and hep-th.}

{The comparison drawn may allow to introduce apt corrections to usual bibliographic indicators, by properly weighting the participation to large collaborations, as opposed to research activities carried out in small groups.  Recall that the hypergraph Laplacian is equivalent to the Laplacian obtained from a properly weighted projected network,  which inherits of the high-order structures of the hypergraph~\cite{chitraraphael2019}. Assessing the higher order ranking amounts therefore to applying the usual tools to this latter weighted binary graph, a conclusion which points to an immediate operative translation of the newly introduced methods. }

\subsection*{Classification task}
\label{sec_app_zoo}

{To further test the interest of a generalised random walk process biased to account for hyperedged communities within a plausible microscopic framework, we consider the classification task studied by Zhou et al.~\cite{zhou2007learning}. We anticipate that the 
obtained classification outperforms that obtained under the usual random walk framework, which ignores the annotated hyper structures.}

{A standard pipeline to analyse a dataset starts with the determination of pairwise similarities between the objects to be eventually classified. This implies defining a network that can be studied by means of standard spectral methods. However, similarities involve often groups of objects. In this respect, hypergraphs define the ideal mathematical  platform to account for the inherent complexity of the classification problem.  More precisely,  one can make use of spectral methods based on the hypergraph Laplace matrix to eventually obtain a classification which effectively accounts for high-order interaction as displayed in the data~\cite{TTTH2015}.}

\begin{figure}[ht]
\centering
\includegraphics[width=.4\textwidth]{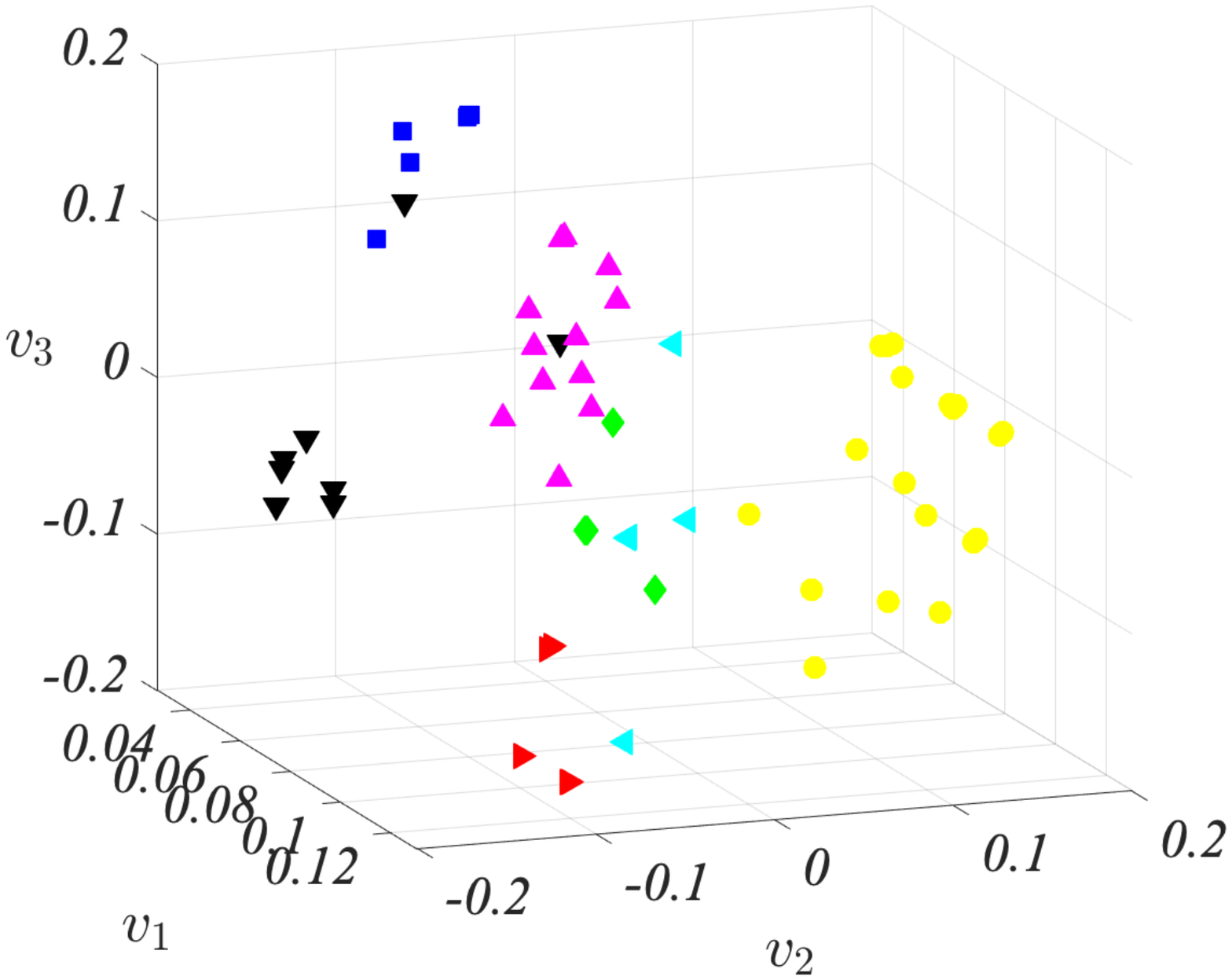}
\vspace{-2cm}
\caption{{\textbf{Classification of the animals according to their features}. We report a $3D$ embedding of the zoo dataset, namely using the first three eigenvectors. Each combination colour/symbol refer to a know class and one can appreciate by eyeball analysis the resulting clusters.}}
\label{fig:zoo}
\end{figure}

{Following ~\cite{zhou2007learning} we consider an ensemble of animals from a zoologically heterogeneous set. Specifically, we used the zoo database taken from the UCI Machine Learning Depository~\cite{Dua:2019}, containing $101$ animals, each one endowed with $16$ features, such as tail, hair, legs and so on. To each animal we associate its corresponding class, e.g. mammals, birds, etc. (see Appendix~\ref{sec:zoo}). Here nodes are animals and hyperedges features; we will show that the presence of high-order interactions among features allow to obtain a very satisfying embedding using only $2$ or $3$ dimensions, a result which is in line with that reported in Zhou et al.~\cite{zhou2007learning} for an {\it ad hoc} choice of the free weights parameters.
To this end we build a hypergraph using the above recipe, we compute its random walk Laplacian and eventually its ensuing spectrum. We list the eigenvalues in ascending order and rename accordingly the eigenvectors. We use the first left eigenvectors~\footnote{In principle also the right eigenvectors can be used for classification purposes.}, associated to the smallest eigenvalues}, as coordinates of a Euclidean space where to embed the data (see Fig.~\ref{fig:zoo}). Let us observe that since we use a random walk Laplacian, the first eigenvector, i.e. the one associated to the $0$ eigenvalue, is not homogeneous and it already contains non trivial information on the structure of the examined sample. Classes are identified by different colours: mammal (yellow circle), bird (magenta up triangle), reptile (cyan left triangle), fish (red right triangle), amphibian (green diamond), bug (blue square) and invertebrate (black down triangle). One can visually appreciate homologous symbols do cluster in space, hence suggesting that the embedding yields an accurate classification. {Indeed, the ground-truth partition of animals into these seven classes, and the one obtained by performing a K-means clustering in this 3-dimensional space have an Adjusted Rand Index (ARI)~\cite{HA1985} equal to $0.54$. In Appendix~\ref{sec:zoo} we show the results that we obtain when using the eigenvectors of the Laplacian of the projected network. In this case the method is less performant. The classification task is hard to be reached with such a small number of dimensions if the hyperedges are not at play, and the found clusters are not correlated to the ground-truth partition, returning a value ARI $=-0.03$.}

%
\section*{Conclusions}	

Summing up, we have here introduced a new class of random walks on hypergraphs which take into account the presence of higher-order interactions. We provided an analytical expression for the ensuing stationary distribution, based on the structural features of the networked system, and compared it to the distribution associated to a traditional random walk performed on the corresponding projected network. {More precisely, we proposed a self-consistent recipe grounded on a microscopic physical random process biased by the hyperedges sizes  to assign weights to hyperedges.} We further characterised the dynamics by comparing the two processes on several synthetic and real-world networks, both by means of numerical simulations and analytical arguments. We show that our process produces {stationary distributions different from those} obtained for the corresponding projected network, and that prove sensitive to higher-order structure in a networked architecture. {Our framework is applied to collaboration networks, yielding new insights on node ranking and centrality measure, which allow for a richer characterisation of individual performances, as compared to traditional methods. Moreover, we show that information embedded in the higher-order walk can be used to achieve accurate classification. In particular, we applied our method to successfully cluster into different families, animals with different features, each one representing an hyperedge. The same procedure fails if a simple random walk on the corresponding projected network is considered.  Importantly, the proposed Laplacian is equivalent to that stemming from a properly tuned weighted network~\cite{chitraraphael2019}. Higher oder rankings and refined classifications could be hence immediately obtained by supplying to conventional tools and analysis schemes the weighted adjacency matrix that characterises the graph with pairwise edges associated to the hypergraph construction. Taken all together, our work sheds new light on dynamical processes on networks which are not limited to pairwise interactions, and on the complex interplay between the structure and dynamics of higher-order interaction networks. Future applications to machine learning based approaches to classification are also envisaged.}

{}

	\addcontentsline{toc}{chapter}{Bibliography}
	\bibliography{bib_HRW}

\begin{thebibliography}{63}%
\makeatletter
\providecommand \@ifxundefined [1]{%
 \@ifx{#1\undefined}
}%
\providecommand \@ifnum [1]{%
 \ifnum #1\expandafter \@firstoftwo
 \else \expandafter \@secondoftwo
 \fi
}%
\providecommand \@ifx [1]{%
 \ifx #1\expandafter \@firstoftwo
 \else \expandafter \@secondoftwo
 \fi
}%
\providecommand \natexlab [1]{#1}%
\providecommand \enquote  [1]{``#1''}%
\providecommand \bibnamefont  [1]{#1}%
\providecommand \bibfnamefont [1]{#1}%
\providecommand \citenamefont [1]{#1}%
\providecommand \href@noop [0]{\@secondoftwo}%
\providecommand \href [0]{\begingroup \@sanitize@url \@href}%
\providecommand \@href[1]{\@@startlink{#1}\@@href}%
\providecommand \@@href[1]{\endgroup#1\@@endlink}%
\providecommand \@sanitize@url [0]{\catcode `\\12\catcode `\$12\catcode
  `\&12\catcode `\#12\catcode `\^12\catcode `\_12\catcode `\%12\relax}%
\providecommand \@@startlink[1]{}%
\providecommand \@@endlink[0]{}%
\providecommand \url  [0]{\begingroup\@sanitize@url \@url }%
\providecommand \@url [1]{\endgroup\@href {#1}{\urlprefix }}%
\providecommand \urlprefix  [0]{URL }%
\providecommand \Eprint [0]{\href }%
\providecommand \doibase [0]{http://dx.doi.org/}%
\providecommand \selectlanguage [0]{\@gobble}%
\providecommand \bibinfo  [0]{\@secondoftwo}%
\providecommand \bibfield  [0]{\@secondoftwo}%
\providecommand \translation [1]{[#1]}%
\providecommand \BibitemOpen [0]{}%
\providecommand \bibitemStop [0]{}%
\providecommand \bibitemNoStop [0]{.\EOS\space}%
\providecommand \EOS [0]{\spacefactor3000\relax}%
\providecommand \BibitemShut  [1]{\csname bibitem#1\endcsname}%
\let\auto@bib@innerbib\@empty
\bibitem [{\citenamefont {Newman}(2010)}]{Newmanbook}%
  \BibitemOpen
  \bibfield  {author} {\bibinfo {author} {\bibfnamefont {Mark~EJ}\ \bibnamefont
  {Newman}},\ }\href@noop {} {\emph {\bibinfo {title} {Networks: An
  Introduction}}}\ (\bibinfo  {publisher} {Oxford University Press},\ \bibinfo
  {address} {Oxford},\ \bibinfo {year} {2010})\BibitemShut {NoStop}%
\bibitem [{\citenamefont {Barab{\'a}si}\ \emph {et~al.}(2016)\citenamefont
  {Barab{\'a}si} \emph {et~al.}}]{Barabasibook}%
  \BibitemOpen
  \bibfield  {author} {\bibinfo {author} {\bibfnamefont
  {Albert-L{\'a}szl{\'o}}\ \bibnamefont {Barab{\'a}si}} \emph {et~al.},\
  }\href@noop {} {\emph {\bibinfo {title} {Network science}}}\ (\bibinfo
  {publisher} {Cambridge university press},\ \bibinfo {year}
  {2016})\BibitemShut {NoStop}%
\bibitem [{\citenamefont {Latora}\ \emph {et~al.}(2017)\citenamefont {Latora},
  \citenamefont {Nicosia},\ and\ \citenamefont {Russo}}]{Latorabook}%
  \BibitemOpen
  \bibfield  {author} {\bibinfo {author} {\bibfnamefont {Vito}\ \bibnamefont
  {Latora}}, \bibinfo {author} {\bibfnamefont {Vincenzo}\ \bibnamefont
  {Nicosia}}, \ and\ \bibinfo {author} {\bibfnamefont {Giovanni}\ \bibnamefont
  {Russo}},\ }\href@noop {} {\emph {\bibinfo {title} {Complex networks:
  principles, methods and applications}}}\ (\bibinfo  {publisher} {Cambridge
  University Press},\ \bibinfo {year} {2017})\BibitemShut {NoStop}%
\bibitem [{\citenamefont {Albert}\ and\ \citenamefont
  {Barab{\'a}si}(2002)}]{AlbertBarabasi}%
  \BibitemOpen
  \bibfield  {author} {\bibinfo {author} {\bibfnamefont {R{\'e}ka}\
  \bibnamefont {Albert}}\ and\ \bibinfo {author} {\bibfnamefont
  {Albert-L{\'a}szl{\'o}}\ \bibnamefont {Barab{\'a}si}},\ }\bibfield  {title}
  {\enquote {\bibinfo {title} {Statistical mechanics of complex networks},}\
  }\href@noop {} {\bibfield  {journal} {\bibinfo  {journal} {Reviews of modern
  physics}\ }\textbf {\bibinfo {volume} {74}},\ \bibinfo {pages} {47} (\bibinfo
  {year} {2002})}\BibitemShut {NoStop}%
\bibitem [{\citenamefont {Boccaletti}\ \emph {et~al.}(2006)\citenamefont
  {Boccaletti}, \citenamefont {Latora}, \citenamefont {Moreno}, \citenamefont
  {Chavez},\ and\ \citenamefont {Hwang}}]{BLMCH}%
  \BibitemOpen
  \bibfield  {author} {\bibinfo {author} {\bibfnamefont {Stefano}\ \bibnamefont
  {Boccaletti}}, \bibinfo {author} {\bibfnamefont {Vito}\ \bibnamefont
  {Latora}}, \bibinfo {author} {\bibfnamefont {Yamir}\ \bibnamefont {Moreno}},
  \bibinfo {author} {\bibfnamefont {Martin}\ \bibnamefont {Chavez}}, \ and\
  \bibinfo {author} {\bibfnamefont {D-U}\ \bibnamefont {Hwang}},\ }\bibfield
  {title} {\enquote {\bibinfo {title} {Complex networks: Structure and
  dynamics},}\ }\href@noop {} {\bibfield  {journal} {\bibinfo  {journal}
  {Physics reports}\ }\textbf {\bibinfo {volume} {424}},\ \bibinfo {pages}
  {175--308} (\bibinfo {year} {2006})}\BibitemShut {NoStop}%
\bibitem [{\citenamefont {Castellano}\ \emph {et~al.}(2009)\citenamefont
  {Castellano}, \citenamefont {Fortunato},\ and\ \citenamefont
  {Loreto}}]{Castellanoreview}%
  \BibitemOpen
  \bibfield  {author} {\bibinfo {author} {\bibfnamefont {Claudio}\ \bibnamefont
  {Castellano}}, \bibinfo {author} {\bibfnamefont {Santo}\ \bibnamefont
  {Fortunato}}, \ and\ \bibinfo {author} {\bibfnamefont {Vittorio}\
  \bibnamefont {Loreto}},\ }\bibfield  {title} {\enquote {\bibinfo {title}
  {Statistical physics of social dynamics},}\ }\href@noop {} {\bibfield
  {journal} {\bibinfo  {journal} {Reviews of modern physics}\ }\textbf
  {\bibinfo {volume} {81}},\ \bibinfo {pages} {591} (\bibinfo {year}
  {2009})}\BibitemShut {NoStop}%
\bibitem [{\citenamefont {Arenas}\ \emph {et~al.}(2008)\citenamefont {Arenas},
  \citenamefont {D{\'\i}az-Guilera}, \citenamefont {Kurths}, \citenamefont
  {Moreno},\ and\ \citenamefont {Zhou}}]{Arenasreview}%
  \BibitemOpen
  \bibfield  {author} {\bibinfo {author} {\bibfnamefont {Alex}\ \bibnamefont
  {Arenas}}, \bibinfo {author} {\bibfnamefont {Albert}\ \bibnamefont
  {D{\'\i}az-Guilera}}, \bibinfo {author} {\bibfnamefont {Jurgen}\ \bibnamefont
  {Kurths}}, \bibinfo {author} {\bibfnamefont {Yamir}\ \bibnamefont {Moreno}},
  \ and\ \bibinfo {author} {\bibfnamefont {Changsong}\ \bibnamefont {Zhou}},\
  }\bibfield  {title} {\enquote {\bibinfo {title} {Synchronization in complex
  networks},}\ }\href@noop {} {\bibfield  {journal} {\bibinfo  {journal}
  {Physics reports}\ }\textbf {\bibinfo {volume} {469}},\ \bibinfo {pages}
  {93--153} (\bibinfo {year} {2008})}\BibitemShut {NoStop}%
\bibitem [{\citenamefont {Lambiotte}\ \emph {et~al.}(2019)\citenamefont
  {Lambiotte}, \citenamefont {Rosvall},\ and\ \citenamefont {Scholtes}}]{LRS}%
  \BibitemOpen
  \bibfield  {author} {\bibinfo {author} {\bibfnamefont {Renaud}\ \bibnamefont
  {Lambiotte}}, \bibinfo {author} {\bibfnamefont {Martin}\ \bibnamefont
  {Rosvall}}, \ and\ \bibinfo {author} {\bibfnamefont {Ingo}\ \bibnamefont
  {Scholtes}},\ }\bibfield  {title} {\enquote {\bibinfo {title} {From networks
  to optimal higher-order models of complex systems},}\ }\href@noop {}
  {\bibfield  {journal} {\bibinfo  {journal} {Nature physics}\ ,\ \bibinfo
  {pages} {1}} (\bibinfo {year} {2019})}\BibitemShut {NoStop}%
\bibitem [{\citenamefont {Scholtes}\ \emph {et~al.}(2014)\citenamefont
  {Scholtes}, \citenamefont {Wider}, \citenamefont {Pfitzner}, \citenamefont
  {Garas}, \citenamefont {Tessone},\ and\ \citenamefont
  {Schweitzer}}]{Scholtes}%
  \BibitemOpen
  \bibfield  {author} {\bibinfo {author} {\bibfnamefont {Ingo}\ \bibnamefont
  {Scholtes}}, \bibinfo {author} {\bibfnamefont {Nicolas}\ \bibnamefont
  {Wider}}, \bibinfo {author} {\bibfnamefont {Ren{\'e}}\ \bibnamefont
  {Pfitzner}}, \bibinfo {author} {\bibfnamefont {Antonios}\ \bibnamefont
  {Garas}}, \bibinfo {author} {\bibfnamefont {Claudio~J}\ \bibnamefont
  {Tessone}}, \ and\ \bibinfo {author} {\bibfnamefont {Frank}\ \bibnamefont
  {Schweitzer}},\ }\bibfield  {title} {\enquote {\bibinfo {title}
  {Causality-driven slow-down and speed-up of diffusion in non-markovian
  temporal networks},}\ }\href@noop {} {\bibfield  {journal} {\bibinfo
  {journal} {Nature communications}\ }\textbf {\bibinfo {volume} {5}},\
  \bibinfo {pages} {5024} (\bibinfo {year} {2014})}\BibitemShut {NoStop}%
\bibitem [{\citenamefont {Rosvall}\ \emph {et~al.}(2014)\citenamefont
  {Rosvall}, \citenamefont {Esquivel}, \citenamefont {Lancichinetti},
  \citenamefont {West},\ and\ \citenamefont {Lambiotte}}]{RELWL}%
  \BibitemOpen
  \bibfield  {author} {\bibinfo {author} {\bibfnamefont {Martin}\ \bibnamefont
  {Rosvall}}, \bibinfo {author} {\bibfnamefont {Alcides~V}\ \bibnamefont
  {Esquivel}}, \bibinfo {author} {\bibfnamefont {Andrea}\ \bibnamefont
  {Lancichinetti}}, \bibinfo {author} {\bibfnamefont {Jevin~D}\ \bibnamefont
  {West}}, \ and\ \bibinfo {author} {\bibfnamefont {Renaud}\ \bibnamefont
  {Lambiotte}},\ }\bibfield  {title} {\enquote {\bibinfo {title} {Memory in
  network flows and its effects on spreading dynamics and community
  detection},}\ }\href@noop {} {\bibfield  {journal} {\bibinfo  {journal}
  {Nature communications}\ }\textbf {\bibinfo {volume} {5}},\ \bibinfo {pages}
  {4630} (\bibinfo {year} {2014})}\BibitemShut {NoStop}%
\bibitem [{\citenamefont {De~Domenico}\ \emph {et~al.}(2016)\citenamefont
  {De~Domenico}, \citenamefont {Granell}, \citenamefont {Porter},\ and\
  \citenamefont {Arenas}}]{DDGPA}%
  \BibitemOpen
  \bibfield  {author} {\bibinfo {author} {\bibfnamefont {Manlio}\ \bibnamefont
  {De~Domenico}}, \bibinfo {author} {\bibfnamefont {Clara}\ \bibnamefont
  {Granell}}, \bibinfo {author} {\bibfnamefont {Mason~A}\ \bibnamefont
  {Porter}}, \ and\ \bibinfo {author} {\bibfnamefont {Alex}\ \bibnamefont
  {Arenas}},\ }\bibfield  {title} {\enquote {\bibinfo {title} {The physics of
  spreading processes in multilayer networks},}\ }\href@noop {} {\bibfield
  {journal} {\bibinfo  {journal} {Nature Physics}\ }\textbf {\bibinfo {volume}
  {12}},\ \bibinfo {pages} {901} (\bibinfo {year} {2016})}\BibitemShut
  {NoStop}%
\bibitem [{\citenamefont {Kivel{\"a}}\ \emph {et~al.}(2014)\citenamefont
  {Kivel{\"a}}, \citenamefont {Arenas}, \citenamefont {Barthelemy},
  \citenamefont {Gleeson}, \citenamefont {Moreno},\ and\ \citenamefont
  {Porter}}]{kivela2014multilayer}%
  \BibitemOpen
  \bibfield  {author} {\bibinfo {author} {\bibfnamefont {Mikko}\ \bibnamefont
  {Kivel{\"a}}}, \bibinfo {author} {\bibfnamefont {Alex}\ \bibnamefont
  {Arenas}}, \bibinfo {author} {\bibfnamefont {Marc}\ \bibnamefont
  {Barthelemy}}, \bibinfo {author} {\bibfnamefont {James~P}\ \bibnamefont
  {Gleeson}}, \bibinfo {author} {\bibfnamefont {Yamir}\ \bibnamefont {Moreno}},
  \ and\ \bibinfo {author} {\bibfnamefont {Mason~A}\ \bibnamefont {Porter}},\
  }\bibfield  {title} {\enquote {\bibinfo {title} {Multilayer networks},}\
  }\href@noop {} {\bibfield  {journal} {\bibinfo  {journal} {Journal of complex
  networks}\ }\textbf {\bibinfo {volume} {2}},\ \bibinfo {pages} {203--271}
  (\bibinfo {year} {2014})}\BibitemShut {NoStop}%
\bibitem [{\citenamefont {Battiston}\ \emph {et~al.}(2017)\citenamefont
  {Battiston}, \citenamefont {Nicosia},\ and\ \citenamefont
  {Latora}}]{Battiston2017challenges}%
  \BibitemOpen
  \bibfield  {author} {\bibinfo {author} {\bibfnamefont {Federico}\
  \bibnamefont {Battiston}}, \bibinfo {author} {\bibfnamefont {Vincenzo}\
  \bibnamefont {Nicosia}}, \ and\ \bibinfo {author} {\bibfnamefont {Vito}\
  \bibnamefont {Latora}},\ }\bibfield  {title} {\enquote {\bibinfo {title} {The
  new challenges of multiplex networks: Measures and models},}\ }\href@noop {}
  {\bibfield  {journal} {\bibinfo  {journal} {The European Physical Journal
  Special Topics}\ }\textbf {\bibinfo {volume} {226}},\ \bibinfo {pages}
  {401--416} (\bibinfo {year} {2017})}\BibitemShut {NoStop}%
\bibitem [{\citenamefont {Benson}\ \emph {et~al.}(2016)\citenamefont {Benson},
  \citenamefont {Gleich},\ and\ \citenamefont {Leskovec}}]{BGL}%
  \BibitemOpen
  \bibfield  {author} {\bibinfo {author} {\bibfnamefont {Austin~R}\
  \bibnamefont {Benson}}, \bibinfo {author} {\bibfnamefont {David~F}\
  \bibnamefont {Gleich}}, \ and\ \bibinfo {author} {\bibfnamefont {Jure}\
  \bibnamefont {Leskovec}},\ }\bibfield  {title} {\enquote {\bibinfo {title}
  {Higher-order organization of complex networks},}\ }\href@noop {} {\bibfield
  {journal} {\bibinfo  {journal} {Science}\ }\textbf {\bibinfo {volume}
  {353}},\ \bibinfo {pages} {163--166} (\bibinfo {year} {2016})}\BibitemShut
  {NoStop}%
\bibitem [{\citenamefont {Benson}\ \emph {et~al.}(2018)\citenamefont {Benson},
  \citenamefont {Abebe}, \citenamefont {Schaub}, \citenamefont {Jadbabaie},\
  and\ \citenamefont {Kleinberg}}]{BASJK}%
  \BibitemOpen
  \bibfield  {author} {\bibinfo {author} {\bibfnamefont {A.~R.}\ \bibnamefont
  {Benson}}, \bibinfo {author} {\bibfnamefont {R.}~\bibnamefont {Abebe}},
  \bibinfo {author} {\bibfnamefont {M.~T.}\ \bibnamefont {Schaub}}, \bibinfo
  {author} {\bibfnamefont {A.}~\bibnamefont {Jadbabaie}}, \ and\ \bibinfo
  {author} {\bibfnamefont {J.}~\bibnamefont {Kleinberg}},\ }\bibfield  {title}
  {\enquote {\bibinfo {title} {Simplicial closure and higher-order link
  prediction},}\ }\href@noop {} {\bibfield  {journal} {\bibinfo  {journal}
  {Proceedings of the National Academy of Sciences}\ }\textbf {\bibinfo
  {volume} {115}},\ \bibinfo {pages} {E11221} (\bibinfo {year}
  {2018})}\BibitemShut {NoStop}%
\bibitem [{\citenamefont {Iacopini}\ \emph {et~al.}(2019)\citenamefont
  {Iacopini}, \citenamefont {Petri}, \citenamefont {Barrat},\ and\
  \citenamefont {Latora}}]{IPBL}%
  \BibitemOpen
  \bibfield  {author} {\bibinfo {author} {\bibfnamefont {Iacopo}\ \bibnamefont
  {Iacopini}}, \bibinfo {author} {\bibfnamefont {Giovanni}\ \bibnamefont
  {Petri}}, \bibinfo {author} {\bibfnamefont {Alain}\ \bibnamefont {Barrat}}, \
  and\ \bibinfo {author} {\bibfnamefont {Vito}\ \bibnamefont {Latora}},\
  }\bibfield  {title} {\enquote {\bibinfo {title} {Simplicial models of social
  contagion},}\ }\href@noop {} {\bibfield  {journal} {\bibinfo  {journal}
  {Nature communications}\ }\textbf {\bibinfo {volume} {10}},\ \bibinfo {pages}
  {2485} (\bibinfo {year} {2019})}\BibitemShut {NoStop}%
\bibitem [{\citenamefont {Grilli}\ \emph {et~al.}(2017)\citenamefont {Grilli},
  \citenamefont {Barab{\'a}s}, \citenamefont {Michalska-Smith},\ and\
  \citenamefont {Allesina}}]{GBMSA}%
  \BibitemOpen
  \bibfield  {author} {\bibinfo {author} {\bibfnamefont {Jacopo}\ \bibnamefont
  {Grilli}}, \bibinfo {author} {\bibfnamefont {Gy{\"o}rgy}\ \bibnamefont
  {Barab{\'a}s}}, \bibinfo {author} {\bibfnamefont {Matthew~J}\ \bibnamefont
  {Michalska-Smith}}, \ and\ \bibinfo {author} {\bibfnamefont {Stefano}\
  \bibnamefont {Allesina}},\ }\bibfield  {title} {\enquote {\bibinfo {title}
  {Higher-order interactions stabilize dynamics in competitive network
  models},}\ }\href@noop {} {\bibfield  {journal} {\bibinfo  {journal}
  {Nature}\ }\textbf {\bibinfo {volume} {548}},\ \bibinfo {pages} {210}
  (\bibinfo {year} {2017})}\BibitemShut {NoStop}%
\bibitem [{\citenamefont {Devriendt}\ and\ \citenamefont
  {Van~Mieghem}(2019)}]{DVVM}%
  \BibitemOpen
  \bibfield  {author} {\bibinfo {author} {\bibfnamefont {Karel}\ \bibnamefont
  {Devriendt}}\ and\ \bibinfo {author} {\bibfnamefont {Piet}\ \bibnamefont
  {Van~Mieghem}},\ }\bibfield  {title} {\enquote {\bibinfo {title} {The simplex
  geometry of graphs},}\ }\href@noop {} {\bibfield  {journal} {\bibinfo
  {journal} {Journal of Complex Networks}\ }\textbf {\bibinfo {volume} {7}},\
  \bibinfo {pages} {469--490} (\bibinfo {year} {2019})}\BibitemShut {NoStop}%
\bibitem [{\citenamefont {Courtney}\ and\ \citenamefont
  {Bianconi}(2016{\natexlab{a}})}]{BC}%
  \BibitemOpen
  \bibfield  {author} {\bibinfo {author} {\bibfnamefont {Owen~T}\ \bibnamefont
  {Courtney}}\ and\ \bibinfo {author} {\bibfnamefont {Ginestra}\ \bibnamefont
  {Bianconi}},\ }\bibfield  {title} {\enquote {\bibinfo {title} {Generalized
  network structures: The configuration model and the canonical ensemble of
  simplicial complexes},}\ }\href@noop {} {\bibfield  {journal} {\bibinfo
  {journal} {Physical Review E}\ }\textbf {\bibinfo {volume} {93}},\ \bibinfo
  {pages} {062311} (\bibinfo {year} {2016}{\natexlab{a}})}\BibitemShut
  {NoStop}%
\bibitem [{\citenamefont {Petri}\ and\ \citenamefont
  {Barrat}(2018{\natexlab{a}})}]{PB}%
  \BibitemOpen
  \bibfield  {author} {\bibinfo {author} {\bibfnamefont {Giovanni}\
  \bibnamefont {Petri}}\ and\ \bibinfo {author} {\bibfnamefont {Alain}\
  \bibnamefont {Barrat}},\ }\bibfield  {title} {\enquote {\bibinfo {title}
  {Simplicial activity driven model},}\ }\href@noop {} {\bibfield  {journal}
  {\bibinfo  {journal} {Physical Review Letters}\ }\textbf {\bibinfo {volume}
  {121}},\ \bibinfo {pages} {228301} (\bibinfo {year}
  {2018}{\natexlab{a}})}\BibitemShut {NoStop}%
\bibitem [{\citenamefont {Berge}(1973)}]{berge1973graphs}%
  \BibitemOpen
  \bibfield  {author} {\bibinfo {author} {\bibfnamefont {Claude}\ \bibnamefont
  {Berge}},\ }\href@noop {} {\emph {\bibinfo {title} {Graphs and
  hypergraphs}}},\ North-Holland Pub. Co.\ (\bibinfo  {publisher} {American
  Elsevier Pub. Co},\ \bibinfo {year} {1973})\BibitemShut {NoStop}%
\bibitem [{\citenamefont {Estrada}\ and\ \citenamefont
  {Rodr{\'\i}guez-Vel{\'a}zquez}(2005)}]{estrada2005complex}%
  \BibitemOpen
  \bibfield  {author} {\bibinfo {author} {\bibfnamefont {Ernesto}\ \bibnamefont
  {Estrada}}\ and\ \bibinfo {author} {\bibfnamefont {Juan~A}\ \bibnamefont
  {Rodr{\'\i}guez-Vel{\'a}zquez}},\ }\bibfield  {title} {\enquote {\bibinfo
  {title} {Complex networks as hypergraphs},}\ }\href@noop {} {\bibfield
  {journal} {\bibinfo  {journal} {arXiv preprint physics/0505137}\ } (\bibinfo
  {year} {2005})}\BibitemShut {NoStop}%
\bibitem [{\citenamefont {Ghoshal}\ \emph {et~al.}(2009)\citenamefont
  {Ghoshal}, \citenamefont {Zlati{\'c}}, \citenamefont {Caldarelli},\ and\
  \citenamefont {Newman}}]{GZCN}%
  \BibitemOpen
  \bibfield  {author} {\bibinfo {author} {\bibfnamefont {Gourab}\ \bibnamefont
  {Ghoshal}}, \bibinfo {author} {\bibfnamefont {Vinko}\ \bibnamefont
  {Zlati{\'c}}}, \bibinfo {author} {\bibfnamefont {Guido}\ \bibnamefont
  {Caldarelli}}, \ and\ \bibinfo {author} {\bibfnamefont {Mark~EJ}\
  \bibnamefont {Newman}},\ }\bibfield  {title} {\enquote {\bibinfo {title}
  {Random hypergraphs and their applications},}\ }\href@noop {} {\bibfield
  {journal} {\bibinfo  {journal} {Physical Review E}\ }\textbf {\bibinfo
  {volume} {79}},\ \bibinfo {pages} {066118} (\bibinfo {year}
  {2009})}\BibitemShut {NoStop}%
\bibitem [{\citenamefont {Petri}\ \emph {et~al.}(2014)\citenamefont {Petri},
  \citenamefont {Expert}, \citenamefont {Turkheimer}, \citenamefont
  {Carhart-Harris}, \citenamefont {Nutt}, \citenamefont {Hellyer},\ and\
  \citenamefont {Vaccarino}}]{petri2014homological}%
  \BibitemOpen
  \bibfield  {author} {\bibinfo {author} {\bibfnamefont {Giovanni}\
  \bibnamefont {Petri}}, \bibinfo {author} {\bibfnamefont {Paul}\ \bibnamefont
  {Expert}}, \bibinfo {author} {\bibfnamefont {Federico}\ \bibnamefont
  {Turkheimer}}, \bibinfo {author} {\bibfnamefont {Robin}\ \bibnamefont
  {Carhart-Harris}}, \bibinfo {author} {\bibfnamefont {David}\ \bibnamefont
  {Nutt}}, \bibinfo {author} {\bibfnamefont {Peter~J}\ \bibnamefont {Hellyer}},
  \ and\ \bibinfo {author} {\bibfnamefont {Francesco}\ \bibnamefont
  {Vaccarino}},\ }\bibfield  {title} {\enquote {\bibinfo {title} {Homological
  scaffolds of brain functional networks},}\ }\href@noop {} {\bibfield
  {journal} {\bibinfo  {journal} {Journal of The Royal Society Interface}\
  }\textbf {\bibinfo {volume} {11}},\ \bibinfo {pages} {20140873} (\bibinfo
  {year} {2014})}\BibitemShut {NoStop}%
\bibitem [{\citenamefont {Petri}\ and\ \citenamefont
  {Barrat}(2018{\natexlab{b}})}]{petri2018simplicial}%
  \BibitemOpen
  \bibfield  {author} {\bibinfo {author} {\bibfnamefont {Giovanni}\
  \bibnamefont {Petri}}\ and\ \bibinfo {author} {\bibfnamefont {Alain}\
  \bibnamefont {Barrat}},\ }\bibfield  {title} {\enquote {\bibinfo {title}
  {Simplicial activity driven model},}\ }\href@noop {} {\bibfield  {journal}
  {\bibinfo  {journal} {Physical review letters}\ }\textbf {\bibinfo {volume}
  {121}},\ \bibinfo {pages} {228301} (\bibinfo {year}
  {2018}{\natexlab{b}})}\BibitemShut {NoStop}%
\bibitem [{\citenamefont {Patania}\ \emph {et~al.}(2017)\citenamefont
  {Patania}, \citenamefont {Petri},\ and\ \citenamefont
  {Vaccarino}}]{patania2017shape}%
  \BibitemOpen
  \bibfield  {author} {\bibinfo {author} {\bibfnamefont {Alice}\ \bibnamefont
  {Patania}}, \bibinfo {author} {\bibfnamefont {Giovanni}\ \bibnamefont
  {Petri}}, \ and\ \bibinfo {author} {\bibfnamefont {Francesco}\ \bibnamefont
  {Vaccarino}},\ }\bibfield  {title} {\enquote {\bibinfo {title} {The shape of
  collaborations},}\ }\href@noop {} {\bibfield  {journal} {\bibinfo  {journal}
  {EPJ Data Science}\ }\textbf {\bibinfo {volume} {6}},\ \bibinfo {pages} {18}
  (\bibinfo {year} {2017})}\BibitemShut {NoStop}%
\bibitem [{\citenamefont {Courtney}\ and\ \citenamefont
  {Bianconi}(2016{\natexlab{b}})}]{courtney2016generalized}%
  \BibitemOpen
  \bibfield  {author} {\bibinfo {author} {\bibfnamefont {Owen~T.}\ \bibnamefont
  {Courtney}}\ and\ \bibinfo {author} {\bibfnamefont {Ginestra}\ \bibnamefont
  {Bianconi}},\ }\bibfield  {title} {\enquote {\bibinfo {title} {Generalized
  network structures: The configuration model and the canonical ensemble of
  simplicial complexes},}\ }\href {\doibase 10.1103/PhysRevE.93.062311}
  {\bibfield  {journal} {\bibinfo  {journal} {Phys. Rev. E}\ }\textbf {\bibinfo
  {volume} {93}},\ \bibinfo {pages} {062311} (\bibinfo {year}
  {2016}{\natexlab{b}})}\BibitemShut {NoStop}%
\bibitem [{\citenamefont {May}(1972)}]{May}%
  \BibitemOpen
  \bibfield  {author} {\bibinfo {author} {\bibfnamefont {Robert~M}\
  \bibnamefont {May}},\ }\bibfield  {title} {\enquote {\bibinfo {title} {Will a
  large complex system be stable?}}\ }\href@noop {} {\bibfield  {journal}
  {\bibinfo  {journal} {Nature}\ }\textbf {\bibinfo {volume} {238}},\ \bibinfo
  {pages} {413} (\bibinfo {year} {1972})}\BibitemShut {NoStop}%
\bibitem [{\citenamefont {Allesina}\ and\ \citenamefont
  {Tang}(2012)}]{Allesina}%
  \BibitemOpen
  \bibfield  {author} {\bibinfo {author} {\bibfnamefont {Stefano}\ \bibnamefont
  {Allesina}}\ and\ \bibinfo {author} {\bibfnamefont {Si}~\bibnamefont
  {Tang}},\ }\bibfield  {title} {\enquote {\bibinfo {title} {Stability criteria
  for complex ecosystems},}\ }\href@noop {} {\bibfield  {journal} {\bibinfo
  {journal} {Nature}\ }\textbf {\bibinfo {volume} {483}},\ \bibinfo {pages}
  {205} (\bibinfo {year} {2012})}\BibitemShut {NoStop}%
\bibitem [{\citenamefont {Pecora}\ and\ \citenamefont
  {Carroll}(1998)}]{Pecora}%
  \BibitemOpen
  \bibfield  {author} {\bibinfo {author} {\bibfnamefont {Louis~M}\ \bibnamefont
  {Pecora}}\ and\ \bibinfo {author} {\bibfnamefont {Thomas~L}\ \bibnamefont
  {Carroll}},\ }\bibfield  {title} {\enquote {\bibinfo {title} {Master
  stability functions for synchronized coupled systems},}\ }\href@noop {}
  {\bibfield  {journal} {\bibinfo  {journal} {Physical review letters}\
  }\textbf {\bibinfo {volume} {80}},\ \bibinfo {pages} {2109} (\bibinfo {year}
  {1998})}\BibitemShut {NoStop}%
\bibitem [{\citenamefont {Redner}(2001)}]{Rednerbook}%
  \BibitemOpen
  \bibfield  {author} {\bibinfo {author} {\bibfnamefont {Sidney}\ \bibnamefont
  {Redner}},\ }\href@noop {} {\emph {\bibinfo {title} {A guide to first-passage
  processes}}}\ (\bibinfo  {publisher} {Cambridge University Press},\ \bibinfo
  {year} {2001})\BibitemShut {NoStop}%
\bibitem [{\citenamefont {Noh}\ and\ \citenamefont {Rieger}(2004)}]{noh2004}%
  \BibitemOpen
  \bibfield  {author} {\bibinfo {author} {\bibfnamefont {Jae~Dong}\
  \bibnamefont {Noh}}\ and\ \bibinfo {author} {\bibfnamefont {Heiko}\
  \bibnamefont {Rieger}},\ }\bibfield  {title} {\enquote {\bibinfo {title}
  {Random walks on complex networks},}\ }\href@noop {} {\bibfield  {journal}
  {\bibinfo  {journal} {Physical review letters}\ }\textbf {\bibinfo {volume}
  {92}},\ \bibinfo {pages} {118701} (\bibinfo {year} {2004})}\BibitemShut
  {NoStop}%
\bibitem [{\citenamefont {Newman}(2005)}]{newman2005}%
  \BibitemOpen
  \bibfield  {author} {\bibinfo {author} {\bibfnamefont {Mark~EJ}\ \bibnamefont
  {Newman}},\ }\bibfield  {title} {\enquote {\bibinfo {title} {A measure of
  betweenness centrality based on random walks},}\ }\href@noop {} {\bibfield
  {journal} {\bibinfo  {journal} {Social networks}\ }\textbf {\bibinfo {volume}
  {27}},\ \bibinfo {pages} {39--54} (\bibinfo {year} {2005})}\BibitemShut
  {NoStop}%
\bibitem [{\citenamefont {Rosvall}\ and\ \citenamefont
  {Bergstrom}(2008)}]{rosvall2008}%
  \BibitemOpen
  \bibfield  {author} {\bibinfo {author} {\bibfnamefont {Martin}\ \bibnamefont
  {Rosvall}}\ and\ \bibinfo {author} {\bibfnamefont {Carl~T}\ \bibnamefont
  {Bergstrom}},\ }\bibfield  {title} {\enquote {\bibinfo {title} {Maps of
  random walks on complex networks reveal community structure},}\ }\href@noop
  {} {\bibfield  {journal} {\bibinfo  {journal} {Proceedings of the National
  Academy of Sciences}\ }\textbf {\bibinfo {volume} {105}},\ \bibinfo {pages}
  {1118--1123} (\bibinfo {year} {2008})}\BibitemShut {NoStop}%
\bibitem [{\citenamefont {Nicosia}\ \emph {et~al.}(2014)\citenamefont
  {Nicosia}, \citenamefont {De~Domenico},\ and\ \citenamefont
  {Latora}}]{nicosia2014}%
  \BibitemOpen
  \bibfield  {author} {\bibinfo {author} {\bibfnamefont {Vincenzo}\
  \bibnamefont {Nicosia}}, \bibinfo {author} {\bibfnamefont {Manlio}\
  \bibnamefont {De~Domenico}}, \ and\ \bibinfo {author} {\bibfnamefont {Vito}\
  \bibnamefont {Latora}},\ }\bibfield  {title} {\enquote {\bibinfo {title}
  {Characteristic exponents of complex networks},}\ }\href@noop {} {\bibfield
  {journal} {\bibinfo  {journal} {EPL (Europhysics Letters)}\ }\textbf
  {\bibinfo {volume} {106}},\ \bibinfo {pages} {58005} (\bibinfo {year}
  {2014})}\BibitemShut {NoStop}%
\bibitem [{\citenamefont {G{\'o}mez-Gardenes}\ and\ \citenamefont
  {Latora}(2008)}]{gardenes2008}%
  \BibitemOpen
  \bibfield  {author} {\bibinfo {author} {\bibfnamefont {Jes{\'u}s}\
  \bibnamefont {G{\'o}mez-Gardenes}}\ and\ \bibinfo {author} {\bibfnamefont
  {Vito}\ \bibnamefont {Latora}},\ }\bibfield  {title} {\enquote {\bibinfo
  {title} {Entropy rate of diffusion processes on complex networks},}\
  }\href@noop {} {\bibfield  {journal} {\bibinfo  {journal} {Physical Review
  E}\ }\textbf {\bibinfo {volume} {78}},\ \bibinfo {pages} {065102} (\bibinfo
  {year} {2008})}\BibitemShut {NoStop}%
\bibitem [{\citenamefont {Cencetti}\ \emph {et~al.}(2018)\citenamefont
  {Cencetti}, \citenamefont {Battiston}, \citenamefont {Fanelli},\ and\
  \citenamefont {Latora}}]{cencetti}%
  \BibitemOpen
  \bibfield  {author} {\bibinfo {author} {\bibfnamefont {Giulia}\ \bibnamefont
  {Cencetti}}, \bibinfo {author} {\bibfnamefont {Federico}\ \bibnamefont
  {Battiston}}, \bibinfo {author} {\bibfnamefont {Duccio}\ \bibnamefont
  {Fanelli}}, \ and\ \bibinfo {author} {\bibfnamefont {Vito}\ \bibnamefont
  {Latora}},\ }\bibfield  {title} {\enquote {\bibinfo {title} {Reactive random
  walkers on complex networks},}\ }\href@noop {} {\bibfield  {journal}
  {\bibinfo  {journal} {Physical Review E}\ }\textbf {\bibinfo {volume} {98}},\
  \bibinfo {pages} {052302} (\bibinfo {year} {2018})}\BibitemShut {NoStop}%
\bibitem [{\citenamefont {Skardal}\ and\ \citenamefont
  {Adhikari}(2018)}]{Skardal}%
  \BibitemOpen
  \bibfield  {author} {\bibinfo {author} {\bibfnamefont {Per~Sebastian}\
  \bibnamefont {Skardal}}\ and\ \bibinfo {author} {\bibfnamefont {Sabina}\
  \bibnamefont {Adhikari}},\ }\bibfield  {title} {\enquote {\bibinfo {title}
  {Dynamics of nonlinear random walks on complex networks},}\ }\href@noop {}
  {\bibfield  {journal} {\bibinfo  {journal} {Journal of Nonlinear Science}\ ,\
  \bibinfo {pages} {1--26}} (\bibinfo {year} {2018})}\BibitemShut {NoStop}%
\bibitem [{\citenamefont {Asllani}\ \emph {et~al.}(2018)\citenamefont
  {Asllani}, \citenamefont {Carletti}, \citenamefont {Di~Patti}, \citenamefont
  {Fanelli},\ and\ \citenamefont {Piazza}}]{Asllani2018}%
  \BibitemOpen
  \bibfield  {author} {\bibinfo {author} {\bibfnamefont {Malbor}\ \bibnamefont
  {Asllani}}, \bibinfo {author} {\bibfnamefont {Timoteo}\ \bibnamefont
  {Carletti}}, \bibinfo {author} {\bibfnamefont {Francesca}\ \bibnamefont
  {Di~Patti}}, \bibinfo {author} {\bibfnamefont {Duccio}\ \bibnamefont
  {Fanelli}}, \ and\ \bibinfo {author} {\bibfnamefont {Francesco}\ \bibnamefont
  {Piazza}},\ }\bibfield  {title} {\enquote {\bibinfo {title} {Hopping in the
  crowd to unveil network topology},}\ }\href@noop {} {\bibfield  {journal}
  {\bibinfo  {journal} {Physical review letters}\ }\textbf {\bibinfo {volume}
  {120}},\ \bibinfo {pages} {158301} (\bibinfo {year} {2018})}\BibitemShut
  {NoStop}%
\bibitem [{\citenamefont {Starnini}\ \emph {et~al.}(2012)\citenamefont
  {Starnini}, \citenamefont {Baronchelli}, \citenamefont {Barrat},\ and\
  \citenamefont {Pastor-Satorras}}]{starnini}%
  \BibitemOpen
  \bibfield  {author} {\bibinfo {author} {\bibfnamefont {Michele}\ \bibnamefont
  {Starnini}}, \bibinfo {author} {\bibfnamefont {Andrea}\ \bibnamefont
  {Baronchelli}}, \bibinfo {author} {\bibfnamefont {Alain}\ \bibnamefont
  {Barrat}}, \ and\ \bibinfo {author} {\bibfnamefont {Romualdo}\ \bibnamefont
  {Pastor-Satorras}},\ }\bibfield  {title} {\enquote {\bibinfo {title} {Random
  walks on temporal networks},}\ }\href@noop {} {\bibfield  {journal} {\bibinfo
   {journal} {Physical Review E}\ }\textbf {\bibinfo {volume} {85}},\ \bibinfo
  {pages} {056115} (\bibinfo {year} {2012})}\BibitemShut {NoStop}%
\bibitem [{\citenamefont {Petit}\ \emph {et~al.}(2018)\citenamefont {Petit},
  \citenamefont {Gueuning}, \citenamefont {Carletti}, \citenamefont {Lauwens},\
  and\ \citenamefont {Lambiotte}}]{Petit2018}%
  \BibitemOpen
  \bibfield  {author} {\bibinfo {author} {\bibfnamefont {Julien}\ \bibnamefont
  {Petit}}, \bibinfo {author} {\bibfnamefont {Martin}\ \bibnamefont
  {Gueuning}}, \bibinfo {author} {\bibfnamefont {Timoteo}\ \bibnamefont
  {Carletti}}, \bibinfo {author} {\bibfnamefont {Ben}\ \bibnamefont {Lauwens}},
  \ and\ \bibinfo {author} {\bibfnamefont {Renaud}\ \bibnamefont {Lambiotte}},\
  }\bibfield  {title} {\enquote {\bibinfo {title} {Random walk on temporal
  networks with lasting edges},}\ }\href@noop {} {\bibfield  {journal}
  {\bibinfo  {journal} {Physical Review E}\ }\textbf {\bibinfo {volume} {98}},\
  \bibinfo {pages} {052307} (\bibinfo {year} {2018})}\BibitemShut {NoStop}%
\bibitem [{\citenamefont {Petit}\ \emph {et~al.}(2019)\citenamefont {Petit},
  \citenamefont {Lambiotte},\ and\ \citenamefont {Carletti}}]{Petit2019}%
  \BibitemOpen
  \bibfield  {author} {\bibinfo {author} {\bibfnamefont {Julien}\ \bibnamefont
  {Petit}}, \bibinfo {author} {\bibfnamefont {Renaud}\ \bibnamefont
  {Lambiotte}}, \ and\ \bibinfo {author} {\bibfnamefont {Timoteo}\ \bibnamefont
  {Carletti}},\ }\bibfield  {title} {\enquote {\bibinfo {title} {Classes of
  random walks on temporal networks},}\ }\href@noop {} {\bibfield  {journal}
  {\bibinfo  {journal} {arXiv preprint arXiv:1903.07453}\ } (\bibinfo {year}
  {2019})}\BibitemShut {NoStop}%
\bibitem [{\citenamefont {De~Domenico}\ \emph {et~al.}(2014)\citenamefont
  {De~Domenico}, \citenamefont {Sol{\'e}-Ribalta}, \citenamefont {G{\'o}mez},\
  and\ \citenamefont {Arenas}}]{dedomenico2014navigability}%
  \BibitemOpen
  \bibfield  {author} {\bibinfo {author} {\bibfnamefont {Manlio}\ \bibnamefont
  {De~Domenico}}, \bibinfo {author} {\bibfnamefont {Albert}\ \bibnamefont
  {Sol{\'e}-Ribalta}}, \bibinfo {author} {\bibfnamefont {Sergio}\ \bibnamefont
  {G{\'o}mez}}, \ and\ \bibinfo {author} {\bibfnamefont {Alex}\ \bibnamefont
  {Arenas}},\ }\bibfield  {title} {\enquote {\bibinfo {title} {Navigability of
  interconnected networks under random failures},}\ }\href@noop {} {\bibfield
  {journal} {\bibinfo  {journal} {Proceedings of the National Academy of
  Sciences}\ }\textbf {\bibinfo {volume} {111}},\ \bibinfo {pages} {8351--8356}
  (\bibinfo {year} {2014})}\BibitemShut {NoStop}%
\bibitem [{\citenamefont {Battiston}\ \emph {et~al.}(2016)\citenamefont
  {Battiston}, \citenamefont {Nicosia},\ and\ \citenamefont
  {Latora}}]{battiston2016}%
  \BibitemOpen
  \bibfield  {author} {\bibinfo {author} {\bibfnamefont {Federico}\
  \bibnamefont {Battiston}}, \bibinfo {author} {\bibfnamefont {Vincenzo}\
  \bibnamefont {Nicosia}}, \ and\ \bibinfo {author} {\bibfnamefont {Vito}\
  \bibnamefont {Latora}},\ }\bibfield  {title} {\enquote {\bibinfo {title}
  {Efficient exploration of multiplex networks},}\ }\href@noop {} {\bibfield
  {journal} {\bibinfo  {journal} {New Journal of Physics}\ }\textbf {\bibinfo
  {volume} {18}},\ \bibinfo {pages} {043035} (\bibinfo {year}
  {2016})}\BibitemShut {NoStop}%
\bibitem [{\citenamefont {Schaub}\ \emph {et~al.}(2018)\citenamefont {Schaub},
  \citenamefont {Benson}, \citenamefont {Horn}, \citenamefont {Lippner},\ and\
  \citenamefont {Jadbabaie}}]{SBHLJ}%
  \BibitemOpen
  \bibfield  {author} {\bibinfo {author} {\bibfnamefont {Michael~T}\
  \bibnamefont {Schaub}}, \bibinfo {author} {\bibfnamefont {Austin~R}\
  \bibnamefont {Benson}}, \bibinfo {author} {\bibfnamefont {Paul}\ \bibnamefont
  {Horn}}, \bibinfo {author} {\bibfnamefont {Gabor}\ \bibnamefont {Lippner}}, \
  and\ \bibinfo {author} {\bibfnamefont {Ali}\ \bibnamefont {Jadbabaie}},\
  }\bibfield  {title} {\enquote {\bibinfo {title} {Random walks on simplicial
  complexes and the normalized hodge laplacian},}\ }\href@noop {} {\bibfield
  {journal} {\bibinfo  {journal} {arXiv preprint arXiv:1807.05044}\ } (\bibinfo
  {year} {2018})}\BibitemShut {NoStop}%
\bibitem [{\citenamefont {Lu}\ and\ \citenamefont {Peng}(2011)}]{lu2011high}%
  \BibitemOpen
  \bibfield  {author} {\bibinfo {author} {\bibfnamefont {Linyuan}\ \bibnamefont
  {Lu}}\ and\ \bibinfo {author} {\bibfnamefont {Xing}\ \bibnamefont {Peng}},\
  }\bibfield  {title} {\enquote {\bibinfo {title} {High-ordered random walks
  and generalized laplacians on hypergraphs},}\ }in\ \href@noop {} {\emph
  {\bibinfo {booktitle} {International Workshop on Algorithms and Models for
  the Web-Graph}}}\ (\bibinfo {organization} {Springer},\ \bibinfo {year}
  {2011})\ pp.\ \bibinfo {pages} {14--25}\BibitemShut {NoStop}%
\bibitem [{\citenamefont {Helali}\ and\ \citenamefont
  {L{\"o}we}(2019)}]{HL2019}%
  \BibitemOpen
  \bibfield  {author} {\bibinfo {author} {\bibfnamefont {Amine}\ \bibnamefont
  {Helali}}\ and\ \bibinfo {author} {\bibfnamefont {Matthias}\ \bibnamefont
  {L{\"o}we}},\ }\bibfield  {title} {\enquote {\bibinfo {title} {Hitting times,
  commute times, and cover times for random walks on random hypergraphs},}\
  }\href@noop {} {\bibfield  {journal} {\bibinfo  {journal} {Statistics and
  Probability Letters}\ }\textbf {\bibinfo {volume} {154}},\ \bibinfo {pages}
  {1--6} (\bibinfo {year} {2019})}\BibitemShut {NoStop}%
\bibitem [{\citenamefont {Zhou}\ \emph {et~al.}(2007)\citenamefont {Zhou},
  \citenamefont {Huang},\ and\ \citenamefont
  {Sch{\"o}lkopf}}]{zhou2007learning}%
  \BibitemOpen
  \bibfield  {author} {\bibinfo {author} {\bibfnamefont {Dengyong}\
  \bibnamefont {Zhou}}, \bibinfo {author} {\bibfnamefont {Jiayuan}\
  \bibnamefont {Huang}}, \ and\ \bibinfo {author} {\bibfnamefont {Bernhard}\
  \bibnamefont {Sch{\"o}lkopf}},\ }\bibfield  {title} {\enquote {\bibinfo
  {title} {Learning with hypergraphs: Clustering, classification, and
  embedding},}\ }in\ \href@noop {} {\emph {\bibinfo {booktitle} {Advances in
  neural information processing systems}}}\ (\bibinfo {year} {2007})\ pp.\
  \bibinfo {pages} {1601--1608}\BibitemShut {NoStop}%
\bibitem [{\citenamefont {Matamalas}\ \emph {et~al.}(2019)\citenamefont
  {Matamalas}, \citenamefont {G\'omez},\ and\ \citenamefont
  {Arenas}}]{MGA2019}%
  \BibitemOpen
  \bibfield  {author} {\bibinfo {author} {\bibfnamefont {Joan~T.}\ \bibnamefont
  {Matamalas}}, \bibinfo {author} {\bibfnamefont {Sergio}\ \bibnamefont
  {G\'omez}}, \ and\ \bibinfo {author} {\bibfnamefont {Alex}\ \bibnamefont
  {Arenas}},\ }\bibfield  {title} {\enquote {\bibinfo {title} {Abrupt phase
  transition of epidemic spreading in simplicial complexes},}\ }\href@noop {}
  {\bibfield  {journal} {\bibinfo  {journal} {arXiv preprint:1910.03069v1
  [physics.soc-ph]}\ } (\bibinfo {year} {2019})}\BibitemShut {NoStop}%
\bibitem [{\citenamefont {Jhun}\ \emph {et~al.}(2019)\citenamefont {Jhun},
  \citenamefont {Jo},\ and\ \citenamefont {Kahng}}]{JJK2019}%
  \BibitemOpen
  \bibfield  {author} {\bibinfo {author} {\bibfnamefont {Bukyoung}\
  \bibnamefont {Jhun}}, \bibinfo {author} {\bibfnamefont {Minjae}\ \bibnamefont
  {Jo}}, \ and\ \bibinfo {author} {\bibfnamefont {B.}~\bibnamefont {Kahng}},\
  }\bibfield  {title} {\enquote {\bibinfo {title} {Simplicial sis model in
  scale-free uniform hypergraph},}\ }\href@noop {} {\bibfield  {journal}
  {\bibinfo  {journal} {arXiv preprint:1910.00375v1 [physics.soc-ph]}\ }
  (\bibinfo {year} {2019})}\BibitemShut {NoStop}%
\bibitem [{\citenamefont {Neuh\"{a}user}\ \emph {et~al.}(2019)\citenamefont
  {Neuh\"{a}user}, \citenamefont {Mellor},\ and\ \citenamefont
  {Lambiotte}}]{NML20192019}%
  \BibitemOpen
  \bibfield  {author} {\bibinfo {author} {\bibfnamefont {Leonie}\ \bibnamefont
  {Neuh\"{a}user}}, \bibinfo {author} {\bibfnamefont {Andrew}\ \bibnamefont
  {Mellor}}, \ and\ \bibinfo {author} {\bibfnamefont {Renaud}\ \bibnamefont
  {Lambiotte}},\ }\bibfield  {title} {\enquote {\bibinfo {title} {Multi-body
  interactions and non-linear consensus dynamics on networked systems},}\
  }\href@noop {} {\bibfield  {journal} {\bibinfo  {journal} {arXiv preprint
  arXiv:1910.09226}\ } (\bibinfo {year} {2019})}\BibitemShut {NoStop}%
\bibitem [{\citenamefont {de~Arruda}\ \emph {et~al.}(2019)\citenamefont
  {de~Arruda}, \citenamefont {Petri},\ and\ \citenamefont
  {Moreno}}]{de2019social}%
  \BibitemOpen
  \bibfield  {author} {\bibinfo {author} {\bibfnamefont {Guilherme~Ferraz}\
  \bibnamefont {de~Arruda}}, \bibinfo {author} {\bibfnamefont {Giovanni}\
  \bibnamefont {Petri}}, \ and\ \bibinfo {author} {\bibfnamefont {Yamir}\
  \bibnamefont {Moreno}},\ }\bibfield  {title} {\enquote {\bibinfo {title}
  {Social contagion models on hypergraphs},}\ }\href@noop {} {\bibfield
  {journal} {\bibinfo  {journal} {arXiv preprint arXiv:1909.11154}\ } (\bibinfo
  {year} {2019})}\BibitemShut {NoStop}%
\bibitem [{\citenamefont {Del~Vicario}\ \emph {et~al.}(2016)\citenamefont
  {Del~Vicario}, \citenamefont {Vivaldo}, \citenamefont {Bessi}, \citenamefont
  {Zollo}, \citenamefont {Scala}, \citenamefont {Caldarelli},\ and\
  \citenamefont {Quattrociocchi}}]{echochambers}%
  \BibitemOpen
  \bibfield  {author} {\bibinfo {author} {\bibfnamefont {Michela}\ \bibnamefont
  {Del~Vicario}}, \bibinfo {author} {\bibfnamefont {Gianna}\ \bibnamefont
  {Vivaldo}}, \bibinfo {author} {\bibfnamefont {Alessandro}\ \bibnamefont
  {Bessi}}, \bibinfo {author} {\bibfnamefont {Fabiana}\ \bibnamefont {Zollo}},
  \bibinfo {author} {\bibfnamefont {Antonio}\ \bibnamefont {Scala}}, \bibinfo
  {author} {\bibfnamefont {Guido}\ \bibnamefont {Caldarelli}}, \ and\ \bibinfo
  {author} {\bibfnamefont {Walter}\ \bibnamefont {Quattrociocchi}},\ }\bibfield
   {title} {\enquote {\bibinfo {title} {Echo chambers: Emotional contagion and
  group polarization on facebook},}\ }\href@noop {} {\bibfield  {journal}
  {\bibinfo  {journal} {Scientific Reports}\ }\textbf {\bibinfo {volume} {6}}
  (\bibinfo {year} {2016})}\BibitemShut {NoStop}%
\bibitem [{\citenamefont {Belkin}\ and\ \citenamefont {Niyogi}(2002)}]{BN2002}%
  \BibitemOpen
  \bibfield  {author} {\bibinfo {author} {\bibfnamefont {M}~\bibnamefont
  {Belkin}}\ and\ \bibinfo {author} {\bibfnamefont {P}~\bibnamefont {Niyogi}},\
  }\bibfield  {title} {\enquote {\bibinfo {title} {Laplacian eigenmaps for
  dimensionality reduction and data representation},}\ }\href@noop {}
  {\bibfield  {journal} {\bibinfo  {journal} {Neural Computation}\ }\textbf
  {\bibinfo {volume} {15}},\ \bibinfo {pages} {1373--1396} (\bibinfo {year}
  {2002})}\BibitemShut {NoStop}%
\bibitem [{Note1()}]{Note1}%
  \BibitemOpen
  \bibinfo {note} {We do not consider here hyperedges with size $1$, because
  they correspond to isolated nodes, i.e. nodes that cannot take part to the
  examined process.}\BibitemShut {Stop}%
\bibitem [{\citenamefont {Chitra}\ and\ \citenamefont
  {Raphael}(2019)}]{chitraraphael2019}%
  \BibitemOpen
  \bibfield  {author} {\bibinfo {author} {\bibfnamefont {Uthsav}\ \bibnamefont
  {Chitra}}\ and\ \bibinfo {author} {\bibfnamefont {Benjamin~J.}\ \bibnamefont
  {Raphael}},\ }\bibfield  {title} {\enquote {\bibinfo {title} {Random walks on
  hypergraphs with edge-dependent vertex weights},}\ }\href@noop {} {\bibfield
  {journal} {\bibinfo  {journal} {arXiv preprint arXiv:1905.08287}\ } (\bibinfo
  {year} {2019})}\BibitemShut {NoStop}%
\bibitem [{\citenamefont {Brin}\ and\ \citenamefont
  {Page}(1998)}]{ilprints361}%
  \BibitemOpen
  \bibfield  {author} {\bibinfo {author} {\bibfnamefont {S.}~\bibnamefont
  {Brin}}\ and\ \bibinfo {author} {\bibfnamefont {L.}~\bibnamefont {Page}},\
  }\bibfield  {title} {\enquote {\bibinfo {title} {The anatomy of a large-scale
  hypertextual web search engine},}\ }in\ \href
  {http://ilpubs.stanford.edu:8090/361/} {\emph {\bibinfo {booktitle} {Seventh
  International World-Wide Web Conference (WWW 1998)}}}\ (\bibinfo {year}
  {1998})\BibitemShut {NoStop}%
\bibitem [{\citenamefont {Page}\ \emph {et~al.}(1999)\citenamefont {Page},
  \citenamefont {Brin}, \citenamefont {Motwani},\ and\ \citenamefont
  {Winograd}}]{ilprints422}%
  \BibitemOpen
  \bibfield  {author} {\bibinfo {author} {\bibfnamefont {Lawrence}\
  \bibnamefont {Page}}, \bibinfo {author} {\bibfnamefont {Sergey}\ \bibnamefont
  {Brin}}, \bibinfo {author} {\bibfnamefont {Rajeev}\ \bibnamefont {Motwani}},
  \ and\ \bibinfo {author} {\bibfnamefont {Terry}\ \bibnamefont {Winograd}},\
  }\href {http://ilpubs.stanford.edu:8090/422/} {\emph {\bibinfo {title} {The
  PageRank Citation Ranking: Bringing Order to the Web.}}},\ \bibinfo {type}
  {Technical Report}\ \bibinfo {number} {1999-66}\ (\bibinfo  {institution}
  {Stanford InfoLab},\ \bibinfo {year} {1999})\ \bibinfo {note} {previous
  number = SIDL-WP-1999-0120}\BibitemShut {NoStop}%
\bibitem [{\citenamefont {Gargiulo}\ \emph {et~al.}(2016)\citenamefont
  {Gargiulo}, \citenamefont {Caen}, \citenamefont {Lambiotte},\ and\
  \citenamefont {Carletti}}]{GargiuloEtAl2016}%
  \BibitemOpen
  \bibfield  {author} {\bibinfo {author} {\bibfnamefont {Floriana}\
  \bibnamefont {Gargiulo}}, \bibinfo {author} {\bibfnamefont {Auguste}\
  \bibnamefont {Caen}}, \bibinfo {author} {\bibfnamefont {Renaud}\ \bibnamefont
  {Lambiotte}}, \ and\ \bibinfo {author} {\bibfnamefont {Timoteo}\ \bibnamefont
  {Carletti}},\ }\bibfield  {title} {\enquote {\bibinfo {title} {The classical
  origin of modern mathematics},}\ }\href@noop {} {\bibfield  {journal}
  {\bibinfo  {journal} {EPJ Data Science}\ }\textbf {\bibinfo {volume} {5}},\
  \bibinfo {pages} {26} (\bibinfo {year} {2016})}\BibitemShut {NoStop}%
\bibitem [{\citenamefont {Tran}\ \emph {et~al.}(462)\citenamefont {Tran},
  \citenamefont {Tran}, \citenamefont {Trang},\ and\ \citenamefont
  {Hieu}}]{TTTH2015}%
  \BibitemOpen
  \bibfield  {author} {\bibinfo {author} {\bibfnamefont {Loc~Hoang}\
  \bibnamefont {Tran}}, \bibinfo {author} {\bibfnamefont {Linh~Hoang}\
  \bibnamefont {Tran}}, \bibinfo {author} {\bibfnamefont {Hoang}\ \bibnamefont
  {Trang}}, \ and\ \bibinfo {author} {\bibfnamefont {Le~Trung}\ \bibnamefont
  {Hieu}},\ }\bibfield  {title} {\enquote {\bibinfo {title} {Combinatorial and
  random walk hypergraph laplacian eigenmaps},}\ }\href@noop {} {\bibfield
  {journal} {\bibinfo  {journal} {International Journal of Machine Learning and
  Computing}\ }\textbf {\bibinfo {volume} {5}},\ \bibinfo {pages} {462}
  (\bibinfo {year} {462})}\BibitemShut {NoStop}%
\bibitem [{\citenamefont {Dua}\ and\ \citenamefont {Graff}(2017)}]{Dua:2019}%
  \BibitemOpen
  \bibfield  {author} {\bibinfo {author} {\bibfnamefont {Dheeru}\ \bibnamefont
  {Dua}}\ and\ \bibinfo {author} {\bibfnamefont {Casey}\ \bibnamefont
  {Graff}},\ }\href {https://archive.ics.uci.edu/ml/datasets/Zoo} {\enquote
  {\bibinfo {title} {{UCI} machine learning repository},}\ } (\bibinfo {year}
  {2017})\BibitemShut {NoStop}%
\bibitem [{Note2()}]{Note2}%
  \BibitemOpen
  \bibinfo {note} {In principle also the right eigenvectors can be used for
  classification purposes.}\BibitemShut {Stop}%
\bibitem [{\citenamefont {Hubert}\ and\ \citenamefont {Arabie}(1985)}]{HA1985}%
  \BibitemOpen
  \bibfield  {author} {\bibinfo {author} {\bibfnamefont {L.}~\bibnamefont
  {Hubert}}\ and\ \bibinfo {author} {\bibfnamefont {P.}~\bibnamefont
  {Arabie}},\ }\bibfield  {title} {\enquote {\bibinfo {title} {Comparing
  partitions},}\ }\href@noop {} {\bibfield  {journal} {\bibinfo  {journal}
  {Journal of Classification}\ }\textbf {\bibinfo {volume} {2}},\ \bibinfo
  {pages} {193--218} (\bibinfo {year} {1985})}\BibitemShut {NoStop}%
\end{thebibliography}%

\onecolumngrid

\section*{Acknowledgements}
T.C. would like to thank Floriana Gargiulo for her help in the data collection process.
This research used resources of the "Plateforme Technologique de Calcul Intensif (PTCI)" 
(http://www.ptci.unamur.be) located at the University of Namur, Belgium, which is supported  by the 
FNRS-FRFC, the Walloon Region, and the University of Namur (Conventions No. 2.5020.11, GEQ U.G006.15, 
1610468 et RW/GEQ2016). The PTCI is member of the "Consortium des \'Equipements de Calcul Intensif  
(C\'ECI)" (http://www.ceci-hpc.be).  


\appendix
\section{About the projected network}
\label{sec:projection}

\begin{figure}[ht]
\centering
\includegraphics[width=.35\textwidth]{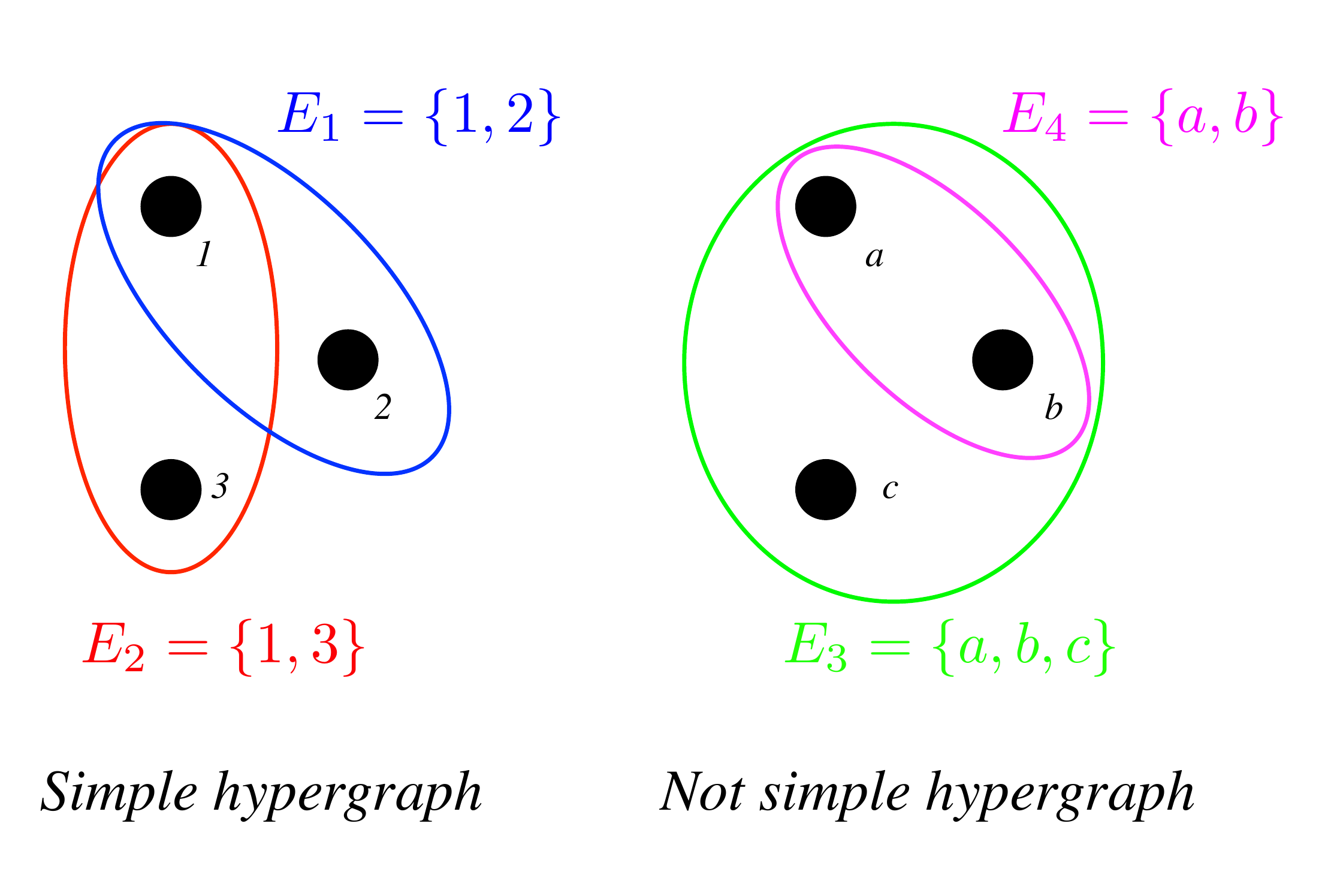}
\caption{\textbf{Simple and not simple hypergraphs}. The hypergraph shown on the left with nodes $V=\{1,2,3\}$ and hyperedges $E_1=\{1,2\}$ and $E_2=\{1,3\}$ is simple, while the one on the right, with nodes $W=\{a,b,c\}$ and hyperedges $E_3=\{a,b,c\}$ and $E_4=\{a,b\}$, is not simple.}
\label{fig:simplenosimple}
\end{figure}

A hypergraph is simple if each hyperedge does not contain any other hyperedge. We report in Fig.~\ref{fig:simplenosimple} two examples, the hypergraph $H_1$ with nodes $V=\{1,2,3\}$ and hyperedges $E_1=\{1,2\}$ and $E_2=\{1,3\}$ is simple because either $E_1\not \subset E_2$ nor $E_2\not \subset E_1$. On the other hand the hypergraph $H_2$ with nodes $W=\{a,b,c\}$ and hyperedges $E_3=\{a,b,c\}$ and $E_4=\{a,b\}$, is not simple because $E_4\subset E_3$. 

\begin{figure}[ht]
\centering
\includegraphics[width=.35\textwidth]{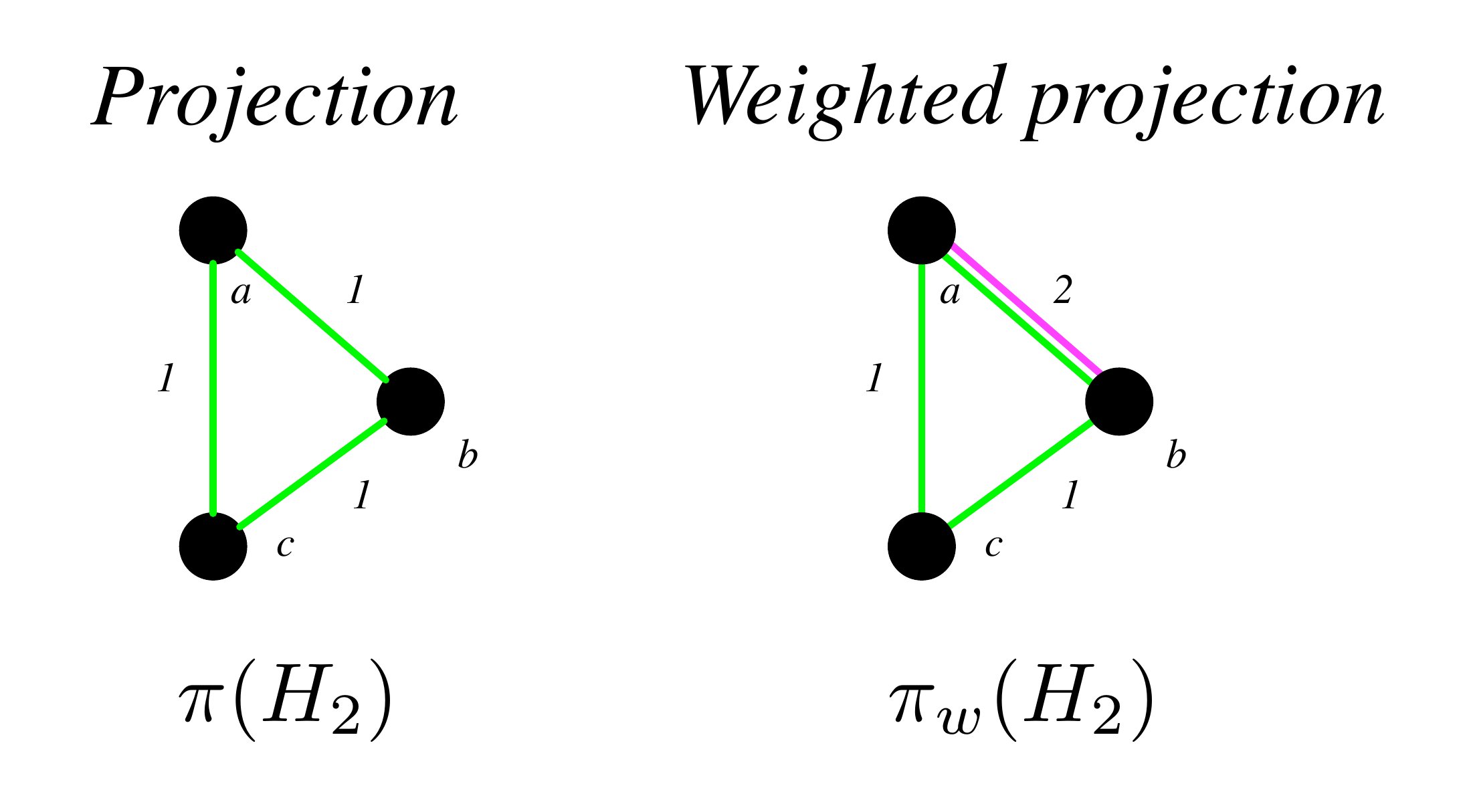}
\caption{\textbf{Weighted projection of hypergraphs}. We propose a standard projection (left panel) and a weighted projection (right panel) of the hypergraph $H_2$ shown in Fig.~\ref{fig:simplenosimple}. In the latter case, the edge $(a,b)$ has weight $2$ because it belongs to two different hyperedges in $H_2$.}
\label{fig:projsimplenosimple}
\end{figure}

Once we build the projected network, $\pi(H_2)$, starting from the latter hypergraph we get a complete $3$-clique, loosing thus information on the existence of  hyperedge $E_4$ (see left panel Fig.~\ref{fig:projsimplenosimple}). Hence, we cannot get back to $H_2$, by inverting the construction, $\pi^{-1} \pi(H_2) \neq H_2$. 
A possible way to overcome this difficulty is to consider a weighted projection (see right panel Fig.~\ref{fig:projsimplenosimple}) where edges inherit a weight counting the number of different hyperedges they belong to. Observe however that for large hyperedge sizes the inversion can be computationally costly because of the combinatorial structure of the problem.

\section{Transition probability}
\label{sec:Tp}
The aim of this section is to provide some details about the calculation of the generalised transition probabilities which take into account the high-order structure of the hyperedges. To compute the transition probability to jump from $i$ to $j$, we  first count the number of nodes, excluding node $i$ itself, belonging to the same hyperedge of $i$ and $j$:
\begin{equation}
\label{eq:dij}
k^{H}_{ij}=\sum_{\alpha}(C_{\alpha \alpha}-1)e_{i\alpha}e_{j\alpha}\quad i\neq j \text{ and }k^{H}_{ii}=0\quad \forall i \, ;
\end{equation}
namely for each hyper edge $E_{\alpha}$ we consider the number of its nodes minus one, i.e. $C_{\alpha \alpha}-1$. Then, this quantity is added to $k^{H}_{ij}$ if and only if $e_{i\alpha}=e_{j\alpha}=1$, that is if and only if both $i$ and $j$ belong to $E_{\alpha}$. 

Secondly, we normalise this quantity by considering a uniform choice among the connected hyperedges. Hence, we obtain a first formula for the transition probability $T_{ij}$ to jump from node $i$ to node $j$:
\begin{equation}
\label{eq:Tij3}
T_{ij}=\frac{k^{H}_{ij}}{\sum_l k^{H}_{il}}=\frac{\sum_{\alpha}(C_{\alpha \alpha}-1)e_{i\alpha}e_{j\alpha}}{\sum_l \sum_{\alpha}(C_{\alpha \alpha}-1)e_{i\alpha}e_{l\alpha}}\,
\end{equation}
so that $\sum_j T_{ij}=1$ $\forall i$.

The latter can be rewritten in an equivalent form, which allows one to draw a comparison with the transition probability for unbiased random walks on networks. Indeed. By recalling the definition of $C_{\alpha\beta}=(e^Te)_{\alpha\beta}=\sum_l e_{\alpha l}^T e_{l \beta}=\sum_l e_{l \alpha} e_{l \beta}$, we get $C_{\alpha\alpha}=\sum_l e_{l \alpha} e_{l \alpha}$ and then
\begin{equation}
\sum_{\alpha}C_{\alpha \alpha}e_{i\alpha}e_{j\alpha}=\sum_{\alpha} e_{i\alpha} C_{\alpha \alpha}e^T_{\alpha j}=(e\hat{C}e^T)_{ij}\, ,
\end{equation}
where $\hat{C}$ is a diagonal matrix: the diagonal of  $\hat{C}$  coincides with that of $C$ and its off-diagonal are identically equal to zero. 
This allows us to rewrite Eq.~\eqref{eq:dij} in a more compact way, Eq.~\eqref{eq:khij} in the main text:
\begin{eqnarray*}
k^{H}_{ij}&=&\sum_{\alpha}(C_{\alpha \alpha}-1)e_{i\alpha}e_{j\alpha}=(e\hat{C}e^T)_{ij}-(ee^T)_{ij}\notag\\
&=&(e\hat{C}e^T)_{ij}-A_{ij}\quad\forall i\neq j\, ,
\end{eqnarray*}
where in the last step we used the definition of the hyper adjacency matrix. We thus eventually get Eq.~\eqref{eq:Tij4} as reported in the main text
\begin{equation*}
T_{ij}=\frac{(e\hat{C}e^T)_{ij}-A_{ij}}{\sum_l k^{H}_{il}}=\frac{(e\hat{C}e^T)_{ij}-A_{ij}}{\sum_l (e\hat{C}e^T)_{il}-k^H_i}\, ,
\end{equation*}
where $k^H_i=\sum_l A_{il}$ is the hyperdegree of the node $i$.

Let us observe that this equation remains valid even for not simple hypergraphs. For instance using again the hypergraph $H_2$ shown in Fig.~\ref{fig:simplenosimple}, where the hyperedge $E_4$ is properly included into $E_3$, we get
\begin{equation*}
k_{ab}=(E_3-1)+(E_4-1)=2+1\text{ and } k_{ac} = E_3-1=2\, ,
\end{equation*}
and thus the following transition probabilities
\begin{equation*}
T_{ab}=\frac{3}{5} \text{ and }T_{ac}=\frac{2}{5} \, ,
\end{equation*}
so the transition from $a$ to $b$ is $1.5$ more probable than to $c$ because $a$ and $b$ share two hyperedges.
Among not simple hypergraphs, one has to account for the fact that hyperedges are repeated several times. The theory here proposed holds true also for weighted hyperedges.

\subsection{Nonlinear transition rates}
\label{ssec:nonlin}

In deriving the transition rates Eq.~\eqref{eq:Tij4}, we assumed that the size of the hyperedge linearly correlates with the probability for the walker to perform a jump, one can of course relax this assumption and introduce nonlinear transition rates. In other words, one can add a bias in~\eqref{eq:dij} in the selection rule for a target node $j$, as operated by a walker sitting on node $i$. For example, one can posit:
\begin{equation}
\label{eq:dijgamma}
k^{(H,\gamma)}_{ij}=\sum_{\alpha}(C_{\alpha \alpha}-1)^\gamma e_{i\alpha}e_{j\alpha}\quad i\neq j \text{ and }k^{(H,\gamma)}_{ii}=0\quad \forall i \, .
\end{equation}
In this way large hyperedges are even more favoured, if $\gamma>0$, while the opposite happens if $\gamma<0$, and we eventually get for the transition probabilities
\begin{equation*}
T^{(\gamma)}_{ij}=\frac{k^{(H,\gamma)}_{ij}}{\sum_l k^{(H,\gamma)}_{il}}\, .
\end{equation*}

Clearly, other choices are possible but exploring further generalisations is left for future investigations.

\section{Stationary solution and ranking}
\label{sec:statsol}
Given the transition probability stored in the matrix $\mathbf{T}$, we can obtain the analytical solution for the stationary state ${\bf p}^{(\infty)}$ defined by  ${\bf p}^{(\infty)}={\bf p}^{(\infty)}\mathbf{T}$. By recalling Eq.~\eqref{eq:statnorm} as reported in the main text
\begin{equation*}
p_j^{(\infty)}=\frac{\sum_l (e\hat{C}e^T)_{jl}-k^H_j}{\sum_{ml}\left[ (e\hat{C}e^T)_{ml}-k^H_m\right]}\,,
\end{equation*}
we can straightforwardly verify that it solves the fixed point equation for the governing dynamics. To this end one needs to plug the above equation into Eq.~\eqref{eq:trans} and recall the definition~\eqref{eq:Tij4} for the $T_{ij}$
\begin{equation}
\sum_j \left(\sum_l (e\hat{C}e^T)_{jl}-k^H_j\right) \left(\frac{(e\hat{C}e^T)_{ji}-A_{ji}}{\sum_l (e\hat{C}e^T)_{jl}-k^H_j}-\delta_{ji}\right)=\sum_j \left[\left((e\hat{C}e^T)_{ji}-A_{ji}\right)-\left(\sum_l (e\hat{C}e^T)_{il}-k^H_i\right)\right]=0\, ,
\end{equation}
where the last step has been obtained by observing that $A_{ij}=A_{ji}$ and $(e\hat{C}e^T)_{ij}=(e\hat{C}e^T)_{ji}$.

As stated in the main text, random walks can be used to rank nodes. This is achieved by evaluating the asymptotic probability to get the walkers on the selected node: the larger the probability, the more central  the node. The analytical expression for $\mathbf{p}^{(\infty)}$ indicates that the ranking provided by the random walkers on the hypergraph is proportional to $d^H_i=\sum_j k_{ij}^H$, while it is well known that the ranking that follows the usual randoms walks on the projected network scales proportionally to the node degree, $k_i$. Two nodes, say $i$ and $j$, are thus ranked differently by the two processes, if $k_i>k_j$ but $d^H_i<d^H_j$. As we will now show, the presence of high-order structures can induce a rankings inversion.

\begin{figure}[ht]
\centering
\includegraphics[width=.8\textwidth]{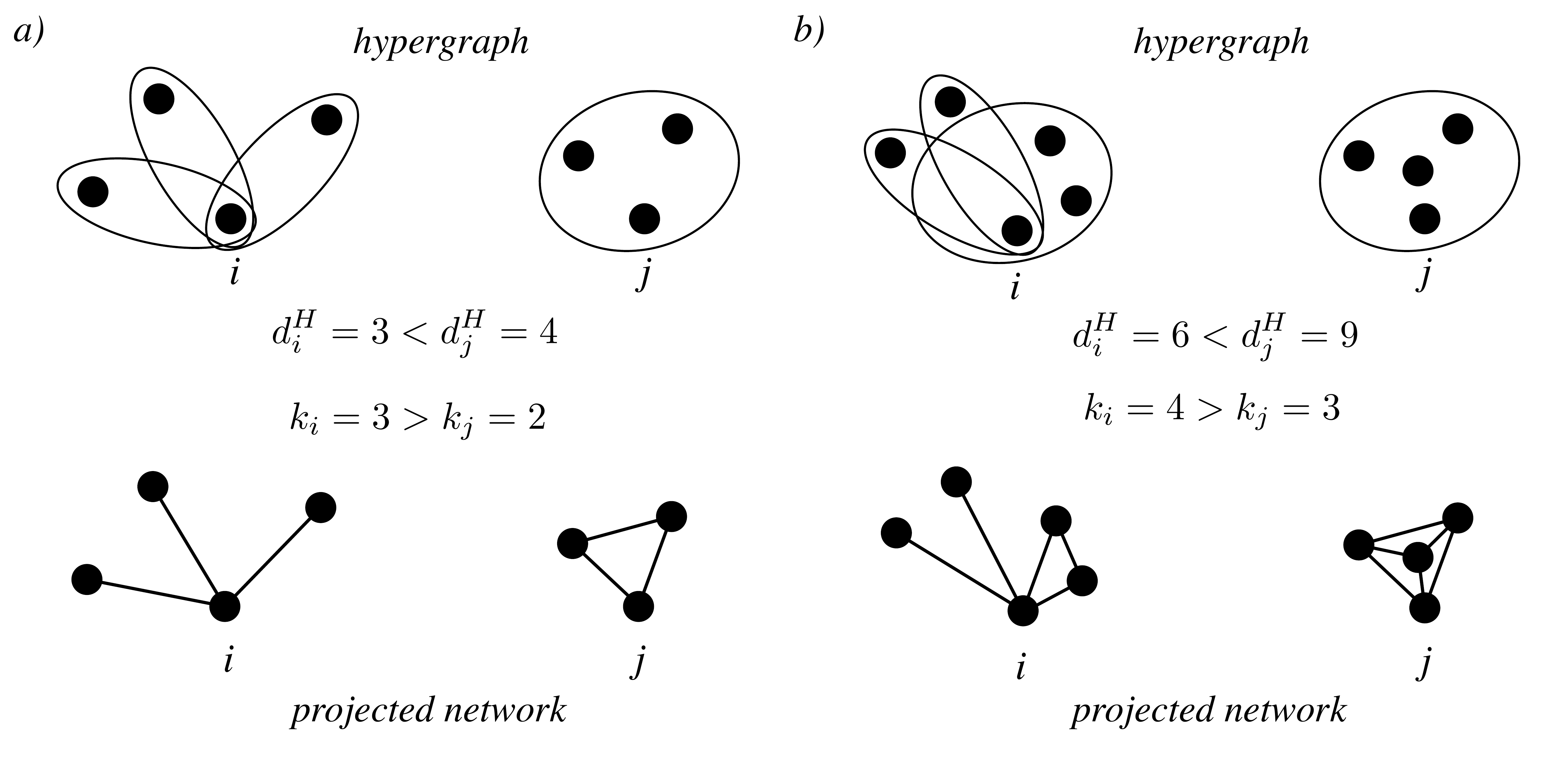}
\caption{{{\bf Examples with ranking inversion.} We propose two typical examples of high-order structures that locally produce two different rankings. On the left panel an example involving three $2$-hyperedges and one $3$-hyperedge, while on the right panel the case with one $3$-hyperedge and two $2$-hyperedges compared with a $4$-hyperedge. In both cases the first configuration will be ranked above the second one, when using the random walks on hypergraphs, while the opposite holds when the random walkers run on the projected networks.}}
\label{fig:rankinv}
\end{figure}

A simple example where this occurs is shown in the left panel of Fig.~\ref{fig:rankinv}. The node $i$ belongs to the intersection of three $2$-hyperedges. Thus its degree (in the projected network) is given by $k_i=3$. Moreover, $d^H_i=3$, because, locally, the hypergraph reduces to a standard network; on the other hand the node $j$ belongs to a $3$-hyperedge, hence $k_j=2$, because it is part of a $3$-clique, but $d^H_j=4$. Hence, $k_i>k_j$ but $d^H_i<d^H_j$. Nodes $j$ will be consequently ranked above $i$ using the generalised random walks on the hypergraph, while the contrary happens if one relies on random walks on the projected network.

The above construction can be readily generalised, as shown by the example presented on the right panel of Fig.~\ref{fig:rankinv}. Here, $i$ belongs to a $3$-hyperedge and to two $2$-hyperedges, hence $k_i=4$ and $d^H_i=6$; node $j$ instead belongs to a $4$-hyperedge, thus $k_j=3$ and $d^H_j=9$. So again $k_i>k_j$ while $d^H_i<d^H_j$.

\section{Heterogeneity of stationary solution}
\label{sec:hetero}
The stationary solution that we obtain from random walk on a hypergraph is very different from the one we can get from the corresponding projected network, the first one being more sensitive to the organisation in groups. The heterogeneity of the state, i.e. the difference among the occupation probability of the different nodes at equilibrium can be quantified by making use of the Gini coefficient.

\begin{figure}[ht]
\centering
\includegraphics[width=.4\textwidth]{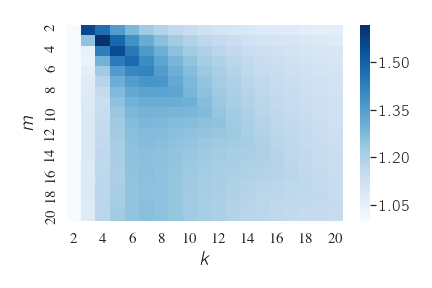}
\caption{{\bf Star-clique model.} Ratio between the Gini coefficient of the stationary state on the hypergraph and the projected network, at varying of the size of the clique ($k$) and of the star ($m$). }
\label{Gini_starclique}
\end{figure}

Fig.~\ref{Gini_starclique} reports on the ratio between the coefficient $G$ computed for the hypergraph and for the projected network of Fig.~\ref{fig:figHnetsEx} of the main text, at varying $m$, the size of the star, and $k$, the size of the clique.

\begin{figure}[ht]
\centering
\includegraphics[width=.4\textwidth]{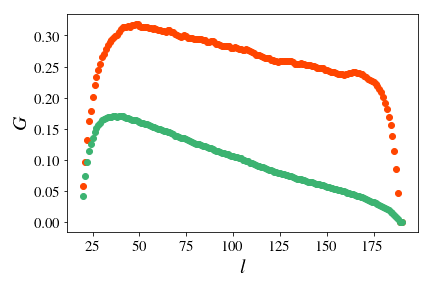}
\caption{{\bf Lattice to fully-connected model} Gini coefficient for the stationary state of the random walk on the hypergraph (red) and on the corresponding projected network (green).}
\label{Gini_1d_to_fc}
\end{figure}

Fig.~\ref{Gini_1d_to_fc} instead shows the heterogeneity for the model which goes from a 1D lattice to a fully connected network, by subsequently adding the links (see Fig.~\ref{fig:RingLinks} of the main text). The red points show the Gini coefficient for the hypergraph, while the green ones are plotted for the projected network, at varying $l$, the number of links in the graph.

{From the results presented in these Figures one can appreciate that the Gini coefficient associated to the stationary solution for the random walk on the hypergraph is always larger than the same quantity computed for the random walk on the projected network. This implies thus that the distribution of walkers on the hypergraph is more heterogeneous than for the projected network.}

\section{Co-authorship networks from arXiv}
\label{sec:arxiv}

The collaboration network is one of the most representative examples of hypergraph; nodes are authors and hyperedges are groups of authors that collaborated to accomplish a task, e.g. write a scientific paper. For this reason we decided to applied the method that we developed to the co-authorship networks extracted from the online preprints platform {\em arXiv} {and hence analyse the nodes ranking obtained using the two processes.}

In this section we report some results for the co-authorship hypergraph for the subdomains of {\em arXiv}, since their existence {up to $2018$ included} (second column Table~\ref{tab:tablearxiv}). In each subdomain, we gathered all the papers and then extracted the authors names, so creating a hyperedge whose nodes are the authors. We thus obtain a set of nodes $V^{(1)}$ and hyperedges $E^{(1)}$, and also the edges of the associated projected network, $E_q^{(1)}$. Such quantities are reported in parentheses in the third, fourth and fifth column of the Table~\ref{tab:tablearxiv}. Once the hypergraph has been built we identify the {\em largest connected component} that will contain the nodes $V^{(cc)}$; then we identify all the maximal, i.e. not properly contained in any other larger hyperedge, and unique hyperedges $E^{(cc)}$ and the edges of the associated projected network $E_q^{(cc)}$. Columns 3-4 and 5 of the Table show such values. Finally, we compute the largest hyperedge and the largest node degree in the maximal connected component (columns 6 and 7). For instance in the arXiv-cs there is a node that belongs to a hyperedge of size $65$ and that is linked to other $406$ nodes: this means that this researcher has signed a paper with $64$ other researchers and in total he/she had $406$ different collaborators with whom he/she has written a paper. Let us also observe that because of the maximality and uniqueness assumptions, we do not know if he/she has co-authored other papers with a subset of the $64$ scholars. Moreover, because we used unweighted networks, we also cannot estimate how many papers he/she wrote with her $406$ collaborators. Let us recall that the need for the maximality and uniqueness is only to compare the results with the projected network, while our method works also without these assumptions.
\begin{table}[h]
\begin{tabular}{l |c|c|c|c|c|c}
\hline
  &   &   &  &   &  &   \\
arXiv & period & nodes & hyperedges& links & $\max |E_{\alpha}| $& $\max k_i$ \\
  &   &   &  &   &  &   \\
\hline
  &   &   &  &   &  &   \\
astro-ph & 1992-2018 & $185579$ ($195729$) & $136918$ ($201270$) & $4602315$ ($4617912$) & $81$ & $2732$ \\
cond-mat & 1992-2018 & $221415$ ($243749$) & $141611$ ($207939$) & $1520895$ ($1551863$) & $63$ & $1426$ \\
cs & 1993-2018 & $136146$ ($187689$) & $84184$ ($139334$) & $534462$ ($607560$) & $65$ & $406$ \\
econ & 2017-2018 & $113$ ($1147$) & $63$ ($612$) & $214$ ($1295$) & $5$ & $36$ \\
gr-qc & 1992-2018 & $32088$ ($40316$) & $25321$ ($45378$) & $216355$ ($228811$) & $80$ & $511$\\
hep-ex & 1992-2018 & $48460$ ($55634$) & $12310$ ($23249$) & $1418268$ ($1435372$) & $83$ & $1228$\\
hep-lat & 1992-2018 & $10275$ ($12483$) & $7439$ ($14143$) & $85194$ ($87254$) & $72$ & $346$\\
hep-ph & 1992-2018 & $62885$ ($70324$) & $50403$ ($86150$) & $814746$ ($823705$) & $74$ & $1244$\\
hep-th & 1992-2018 & $41814$ ($51045$) & $42410$ ($74136$) & $144710$ ($154737$) & $57$ & $206$\\
math & 1992-2018 & $112203$ ($159595$) & $106583$ ($194312$) & $279891$ ($313402$) & $60$ & $336$\\
nlin & 1993-2018 & $19491$ ($30445$) & $12428$ ($23503$) & $52089$ ($64890$) & $46$ & $312$\\
physics & 1996-2018 & $188142$ ($240866$) & $68805$ ($116611$) & $1859156$ ($1950143$) & $80$ & $891$\\
q-bio & 2003-2018 & $23630$ ($45103$) & $9926$ ($21191$) & $93127$ ($142136$) & $54$ & $176$\\
q-fin & 2008-2018 & $3136$ ($8721$) & $2155$ ($6042$) & $6851$ ($13078$) & $11$ & $66$\\
stat & 2008-2018 & $39422$ ($57955$) & $23377$ ($39366$) & $130665$ ($158435$) & $65$ & $228$\\
\hline  
\end{tabular}
\caption{\textbf{Some figures for the arXiv subdomains}. The first column shows the subdomain of the arXiv server, while the second one stands for the period of time for which we have extracted the information. The columns 3, 4 and 5 display respectively the number of nodes, the number of maximal unique hyperedges and the number of links in the largest connected component, while in parenthesis we show the same values for the whole hypergraph/network. In column 6, we report the size of the largest hyperedge and in the 7th  the maximum degree.}
\label{tab:tablearxiv}
\end{table}

\begin{figure}[ht]
\centering
\includegraphics[width=.8\textwidth]{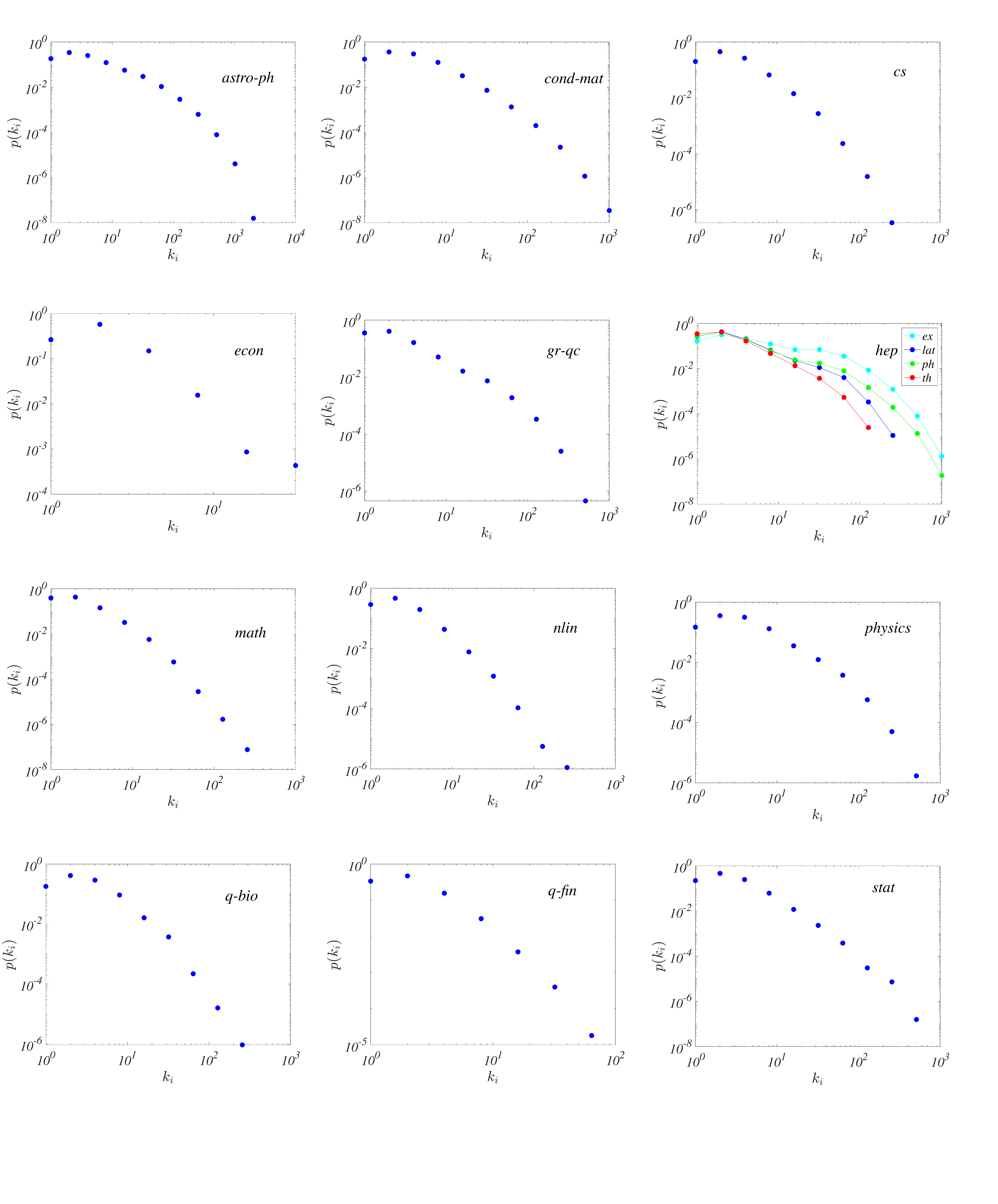}
\caption{\textbf{Degree distribution}. We report for the arXiv subdomains the probability distribution of node degrees, $p(k_i)$, associated to the maximal connected component. In all the cases, we observe a broad distribution; Notice that the arXiv-econ has a relatively small number of papers and authors because of its young age (2017-2018) and thus also the maximal degree, i.e. number of papers written by an author, is quite small.}
\label{fig:figuresarxivdeg}
\end{figure}

\begin{figure}[ht]
\centering
\includegraphics[width=.8\textwidth]{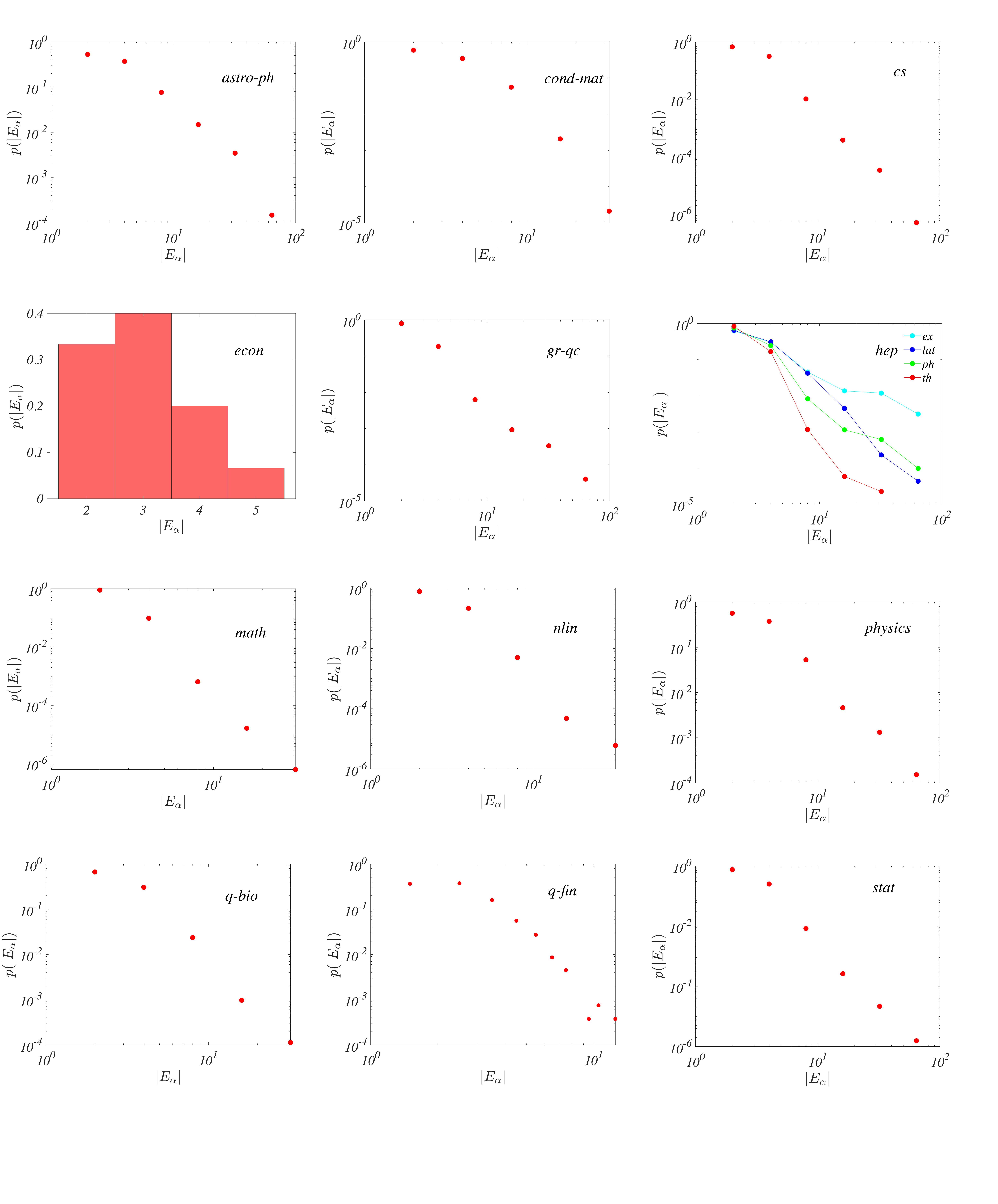}
\caption{\textbf{Hyperedges size distribution}. We report for the arXiv subdomains the probability distribution of hyperedges size, $p(|E_\alpha|)$, associated to the maximal connected component. In all the cases we observe a broad distribution, except for the arXiv-econ for which the number of papers and authors is relatively small because of its young age (2017-2018) and thus also the maximal hyperedge size, i.e. number of co-authors of a paper, is quite small. For this reason we report data in the form of a histogram.}
\label{fig:figuresarxivedge}
\end{figure}

Authors and articles in each subdomain follow different ``rules'' and ``habits'' of publication and writing papers. However, the distribution of node degrees, i.e. number of different collaborators per author, and of hyperedges size, i.e. number of co-authors in papers, exhibit quite similar shapes across the domains, as e.g. broad tails (see Fig.~\ref{fig:figuresarxivdeg} for the degree distribution and - see Fig.~\ref{fig:figuresarxivedge} - for the hyperedges sizes distribution).

\begin{figure}[ht]
\centering
\includegraphics[width=.8\textwidth]{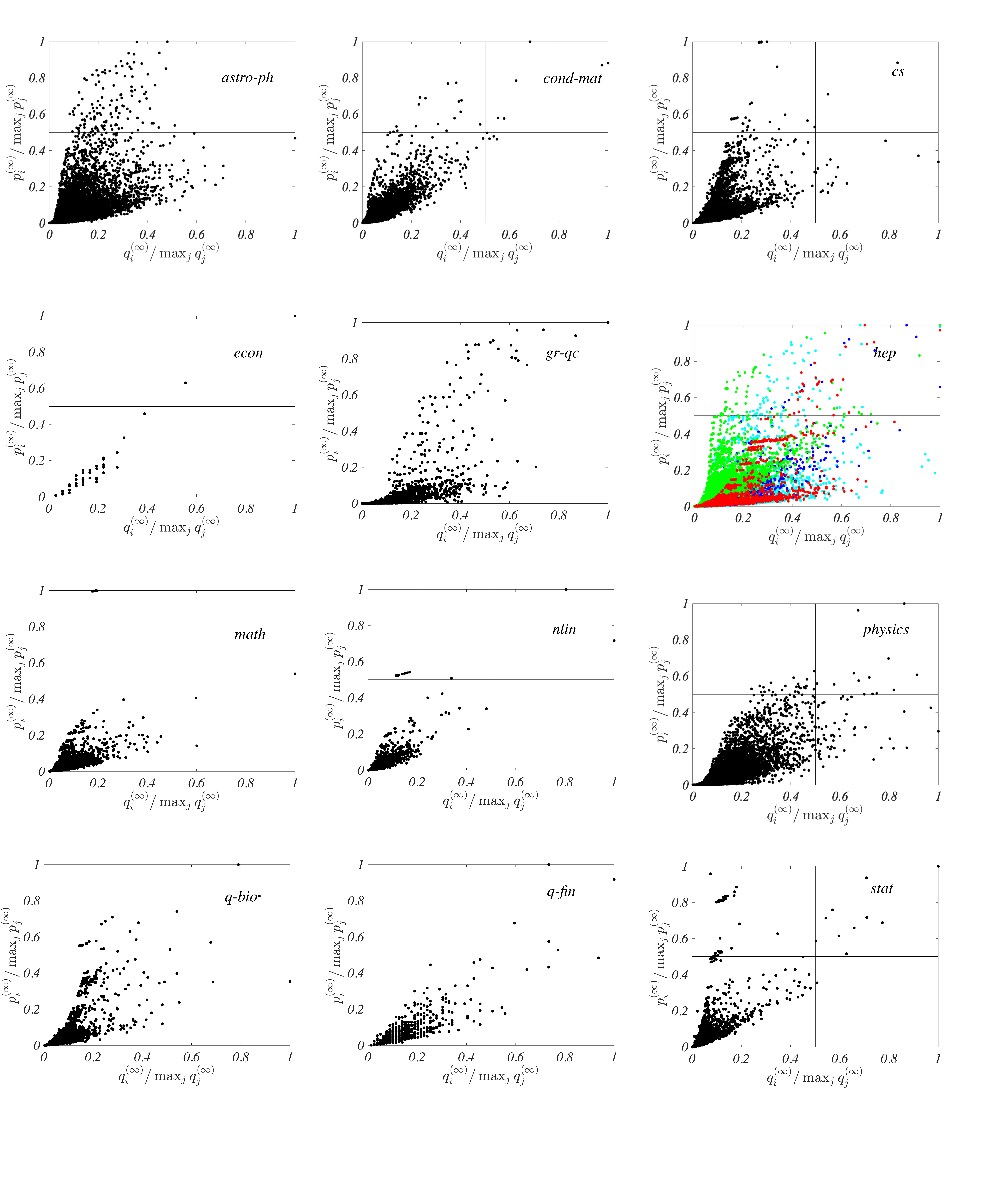}
\caption{\textbf{Comparing the rankings in the arXiv community}. We report the scatter plot of the normalised rankings obtained with the RW on network, $q_i^{(\infty)}$, and the one computed using the random walk on hypergraphs, $p_i^{(\infty)}$.}
\label{fig:arxivQP}
\end{figure}

As already stated the random walk on the hypergraph gives more relevance to the size of the hyperedge, i.e. on the number of co-authors, while the same process on a network emphasises the number of different collaborators. Let us remember that we hereby considered unweighted hypergraphs and networks. We can thus use these approaches to distinguish the different ``publication habits'' in the considered subdomains. To this aim we first normalise the stationary probabilities $p_i^{(\infty)}$ for the hypergraph and $q_i^{(\infty)}$ for the projected network, with respect to their maximum value, to be able to compare sets containing different amount of data, and then we report in the plane with coordinates $\left(q_i^{(\infty)}/\max_j q_j^{(\infty)},p_i^{(\infty)}/\max_jp_j^{(\infty)}\right)$, the scatter plot of the data (each point is an author in the maximal connected component of the hypergraph), separated into different subdomains (see Fig.~\ref{fig:arxivQP}).

If the computed rankings were (almost) the same, the data would (almost) lie on the main diagonal; deviation from this, results in novel information conveyed by the random walk on the hypergraph. Beside the region delimited by $q_i^{(\infty)}/\max_j q_j^{(\infty)} \leq 1/2$ and $p_i^{(\infty)}/\max_jp_j^{(\infty)}\leq 1/2$, associated to authors having written few articles (low degree) and in small group, we identify three interesting zones associated (roughly speaking) to the squares: $[1/2,1]\times [0,1/2]$ (bottom right), $[0,1/2]\times [1/2,1]$ (top left) and $[1/2,1]\times [1/2,1]$ (top right). Authors in the top right square are top ranked in both processes: they have hence written a large number of papers with different collaborators, i.e. large degree, but also they have participated to a relevant number of papers with many co-authors, i.e. large hyperedge size. Scholars in the bottom right square are better ranked by the random walk on the network. This means that they have written several papers but with a small number of co-authors { (see e.g. panel physics in Fig.~\ref{fig:arxivQP})}. Finally scholars in the top left square behave in the opposite way: they have participated to a small number of papers but written by many authors {(see e.g. panels gr-qc, q-bio or stat in Fig.~\ref{fig:arxivQP}).}


\section{The zoo UCI database}
\label{sec:zoo}
{The zoo dataset from the UCI Machine Learning Depository~\cite{Dua:2019}, contains $101$ animals, each one endowed with $15$ boolean features, whose value is thus yes/not, e.g. hair, feathers, eggs, milk, airborne, aquatic, predator, toothed, backbone, breathes, venomous, fins, tail, domestic and catsize. 
There is also a further class that reports on the number of legs, i.e. $0,2,4,6,8$. To homogenise the dataset we decided to introduce five new boolean features to replace the latter one; the new ones being: ``has $0$ legs'', ``has $2$ legs'', ``has $4$ legs'', ``has $6$ legs'', ``has $8$ legs''. The dataset is manually annotated, hence for each animal we have the right class it belongs to, e.g. Mammal, Bird, Reptile, Fish, Amphibian, Bug and Invertebrate.}

{This dataset has been created to provide a benchmark for machine learning tools, to test their capacity to correctly assign each animal to the right class based on the associated features.}

\begin{figure*}[ht]
\centering
\includegraphics[width=.90\textwidth]{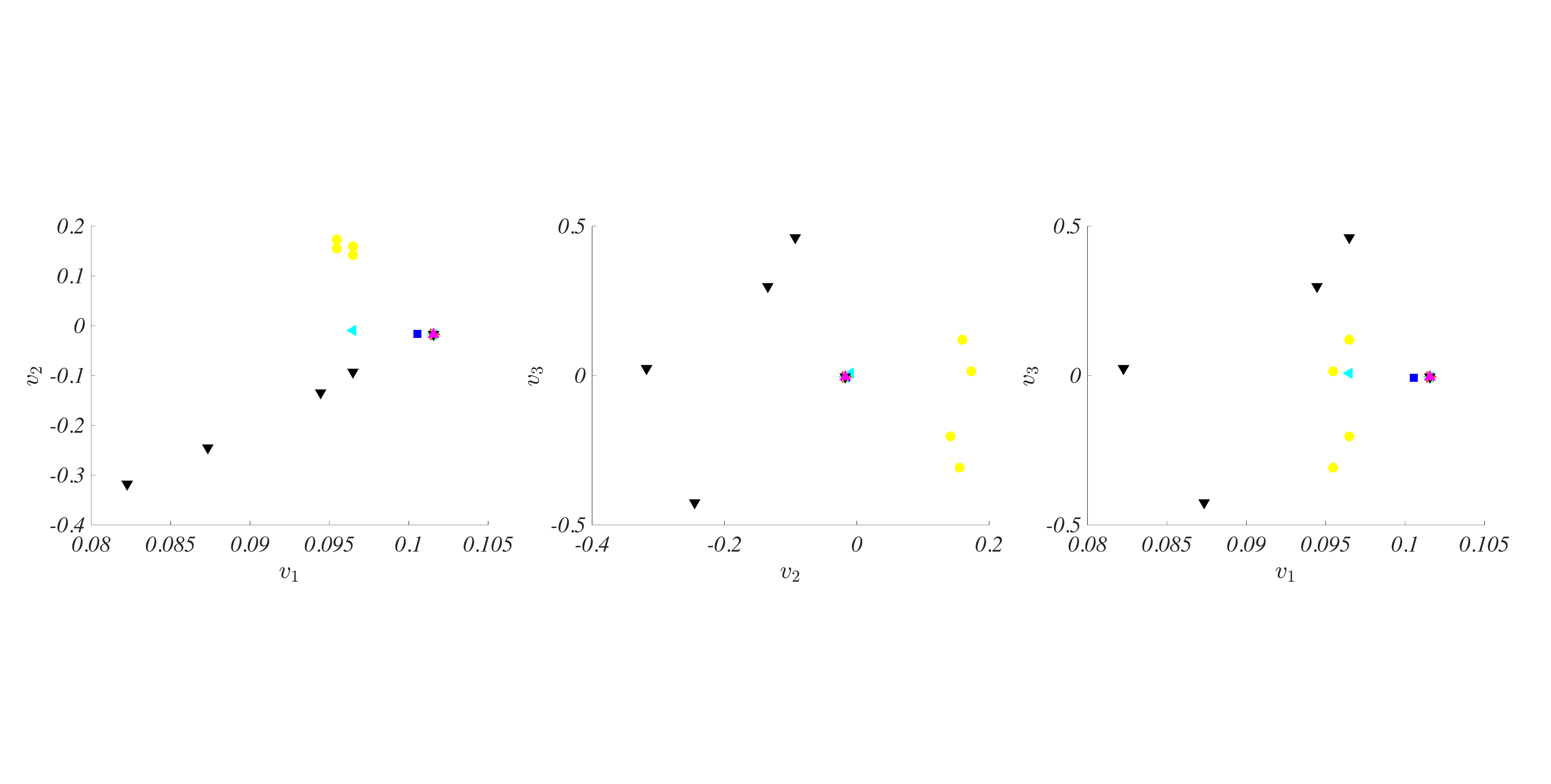}
\vspace{-1cm}
\caption{{\textbf{Classification of the animals according to their features using the projected network}. We report three  $2D$ embedding of the zoo dataset, namely using the first three eigenvector of the random walk Laplacian computed for the projected network. Each combination colour/symbol refer to a know class and one can visually appreciate the resulting clusters.}}
\label{fig:zoo2Dclique}
\end{figure*}

{Animals are the nodes of the hypergraph and the features are the hyperedges, hence all animals sharing the same feature are put in the same hyperedge. The projected network is obtained by making a complete clique from each hyperedge, that is to create a link between all the nodes sharing the same property. Let us observe that this can also be seen as the projection of the bipartite network where there are two kinds of nodes, animals and features, each one linked only to nodes  of the other kind. We thus computed the spectrum of the hypergraph Laplacian and the one for the projected network and we ranked eigenvalues in ascending order, being $0$ the smallest one. We then accordingly rename the associated left eigenvector and we use the first few to embed the data set in small dimensional Euclidean space. Results reported in Fig.~\ref{fig:zoo2Dclique} visually show that classification performances are significantly worse as those obtained when preserving the high order information (see Fig.~\ref{fig:zoo} in the main text). In Fig.~\ref{fig:zoo2Dclique} we report 2D projections, but similar conclusion (as testified by the quantitative ARI scores) are obtained for a 3D embedding of the data. In particular, this method is less sensitive to the differences among nodes and as a consequence, multiple nodes do overlap. Moreover, it is evident that, while some nodes are correctly clustered (like the yellow ones), the others appear confusingly mixed together (see the magenta triangle, which is at the top of a pile of differently coloured symbols).}

%
%
%
%
\end{document}